\documentclass[conference,compsoc]{IEEEtran}
\ifCLASSOPTIONcompsoc
  \usepackage[nocompress]{cite}
\else
  \usepackage{cite}
\fi

\ifCLASSINFOpdf
\else
\fi

\usepackage[T1]{fontenc}
\usepackage{tikz}
\usepackage{amsmath}
\usepackage{amsfonts,amssymb}
\usepackage{algorithm}
\usepackage{algorithmic}
\usepackage{amsthm}
\usepackage{multirow}
\usepackage{graphics}
\usepackage[font=footnotesize]{subfig}
\usepackage{makecell, caption}
\usepackage{booktabs}
\usepackage{framed}
\usepackage{color}
\newtheorem{theorem}{Theorem}

\newcommand{\ie}{\textit{i.e.}}
\newcommand{\eg}{\textit{e.g.}}

\usepackage{amsmath}
\DeclareMathOperator*{\argmin}{argmin}
\usepackage{authblk}

\usepackage{tabulary}

\usepackage{float}

\usepackage[]{footmisc}

\graphicspath{ {fig/} }

\newcommand{\revise}[1]{{#1}}

\begin{document}
%
\title{Prompt Inversion Attack against Collaborative Inference of Large Language Models}

\author{
    Wenjie Qu$^{1,*}$, Yuguang Zhou$^{1,*}$, Yongji Wu$^{2,*}$, Tingsong Xiao$^{3}$,\\
    Binhang Yuan$^{4}$, Yiming Li$^{5}$, Jiaheng Zhang$^{1}$ \\
    { $^{1}$ National University of Singapore, $^{2}$ UC Berkeley, $^{3}$ University of Florida,}\\
    { $^{4}$ Hong Kong University of Science and Technology, $^{5}$ Nanyang Technological University} 
}

\maketitle

\renewcommand{\thefootnote}{}

\footnote{$^*$ The first three authors contributed equally.}


\begin{abstract}
Large language models (LLMs) have been widely applied for their remarkable capability of content generation. However, the practical use of open-source LLMs is hindered by high resource requirements, making deployment expensive and limiting widespread development. The collaborative inference is a promising solution for this problem, in which users collaborate by each hosting a subset of layers and transmitting intermediate activation. Many companies are building collaborative inference platforms to reduce LLM serving costs, leveraging users' underutilized GPUs. Despite widespread interest in collaborative inference within academia and industry, the privacy risks associated with LLM collaborative inference have not been well studied. This is largely because of the challenge posed by inverting LLM activation due to its strong non-linearity.

In this paper, to validate the severity of privacy threats in LLM collaborative inference, we introduce the concept of prompt inversion attack (PIA),  where a malicious participant intends to recover the input prompt through the activation transmitted by its previous participant. Specifically, we design a two-stage method to execute this attack. In the first stage, we optimize the input embedding with a constraint term derived from the LLM's embedding matrix to enforce the optimized embedding to be close to the ground truth. In the second stage, we accurately recover discrete tokens by incorporating activation calibration and semantic speculation.  Extensive experiments show that our PIA method substantially outperforms existing baselines. For example, our method achieves an 88.4\% token accuracy on the Skytrax dataset with the Llama-65B model when inverting the maximum number of transformer layers, while the best baseline method only achieves 22.8\% accuracy. The results verify the effectiveness of our PIA attack and highlights its practical threat to LLM collaborative inference systems.

\end{abstract}

\IEEEpeerreviewmaketitle

\section{Introduction}
\begin{figure}[!t]
\centering
\subfloat{\includegraphics[width=0.97\columnwidth]{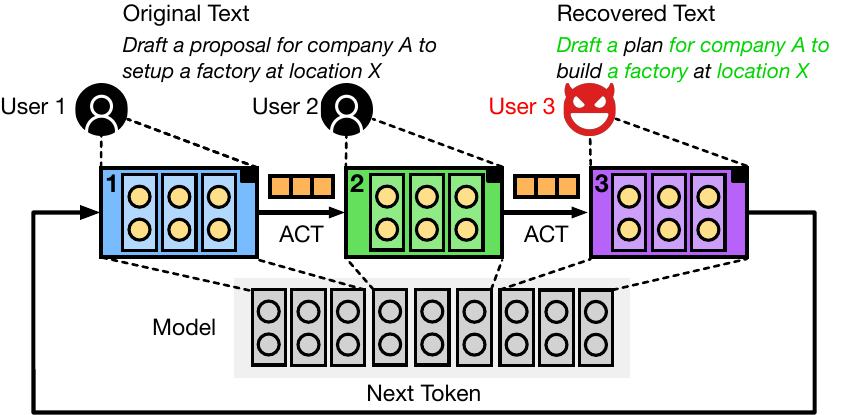}}

\caption{Outline of our attack setting. In collaborative inference, the LLM’s transformer blocks are distributed across collaborating participants, each holding a portion of the model. In this figure, 'ACT' represents the intermediate activation transmitted between participants. User 3 is a malicious participant who records the intermediate activation received from User 2, completing the computation as usual. After inference, User 3 then attempts to reconstruct the input prompt using the recorded activation. }
\label{fig:concept}
\vspace{-1em}
\end{figure}

Large language models (LLMs) (\eg, GPT-4~\cite{gpt4}), have demonstrated exceptional abilities in various domains. The model scale of LLMs has expanded extensively in recent years.
Many LLMs with hundreds of billions of parameters, such as Llama2-70B~\cite{llama2} and OPT-175B~\cite{opt}, have been released.
The practical usage of LLMs is hindered by high memory and computational requirements. For instance, OPT-175B demands over 350 GB of GPU memory for inference. Deploying LLMs necessitates many expensive GPUs, making them unaffordable for most developers. This requirement heavy burden significantly hinders the widespread deployment of LLMs in practice.

Recent studies have been conducted to democratize LLM's access. They can be mainly divided into two categories: offloading and collaborative inference. Offloading-based systems~\cite{sheng2023flexgen,llamacpp,song2023powerinfer} reduce demand on GPU memory by leveraging CPU memory to store model weights that exceed GPU’s capacity. 
However, they suffer from significant serving latency due to frequent transmission of large model weights (\eg, gigabytes in size) between CPU and GPU. Collaborative inference~\cite{PETALS,PETALS2,jiang2023hexgen} is an alternative to offloading, which addresses this challenge. In collaborative inference, participants connect their underutilized GPUs to a pool. 
 Each participant hosts a subset of consecutive LLM layers, receives the preceding layers' activation, computes the hosted layers, and sends the calculated activation to the subsequent participant for follow-up computations. Similar to BitTorrent~\cite{bittorrent}, LLM users benefit from other peoples' GPUs while they can also contribute their own GPUs to the pool. 
The data transmission volume of collaborative inference is typically orders of magnitude smaller than that of offloading, leading to much lower overhead.  Several reports~\cite{speed_petals1,speed_petals2} have pointed out that collaborative inference can output about five tokens per second, similar to human typing speed, which is acceptable in the real world. 

Collaborative inference has recently attracted much attention from industry and academia~\cite{PETALS,yuan2022decentralized,jiang2023hexgen,miao2023sdpipe}.
Multiple companies are building collaborative inference platforms to meet the demand for AI computing power. For example, Huggingface~\cite{Huggingface} and Yandex research~\cite{Yandex} built PETALS~\cite{PETALS,PETALS2}, enabling users to share their underutilized GPUs through the network to offer people access to cheap computing power.   An influential open source project LocalAI~\cite{localai} also added support to collaborative inference to reduce cost and enhance efficiency~\cite{localai_deml}. Meanwhile, the startup Nesa~\cite{NESA} has combined collaborative inference with blockchain technology to improve censorship resistance. In academia, extensive research~\cite{jiang2023hexgen,yuan2022decentralized,miao2023sdpipe,skyserve} has focused on enhancing the efficiency and scalability of collaborative inference. Additionally, several studies~\cite{edgecloud2,edgecloud3,edgecloud1} have explored the specific case of LLM collaborative inference in edge cloud environments.

However, due to the complex nature of the internet, malicious users are likely to exist. For example, a recent work~\cite{luposition} (ICML'24) studied the risk of poisoning under collaborative fine-tuning. In particular, there are two potential threats collaborative inference systems may encounter: (1) \emph{utility threats} that \revise{a malicious client might intentionally delay processing to increase inference latency or alter the activations sent to subsequent participants, thereby manipulating the generated text} and (2) \emph{privacy threats} that infer sensitive information from the activation received from the previous participant.  
This paper mainly focuses on the privacy threats in LLM collaborative inference since input prompts often contain sensitive information (\eg, personal names and contact details) and high-quality LLM prompts themselves hold significant commercial value~\cite{promptmarket}.

To the best of our knowledge, there is no research regarding the privacy threats in collaborative LLM inference. Nevertheless, we notice that some techniques, including attribute inference attack~\cite{pan2020privacy,li2022you,lyu2020differentially} and embedding inversion attack~\cite{song2020information, Generative-embedding-inversion-attack, Sentence_inference_attack}, are closely related to it. These attacks have similar threat models with collaborative inference where the attacker is given embedding/activation. Specifically, attribute inference attacks can accurately infer various sensitive categorized attributes (\eg, location and gender) from language embedding with high accuracy. However, \emph{they are unlikely to practically threaten LLM collaborative inference}, since  LLM input prompts are typically long and often contain multiple different values for a single attribute, making the attack results confusing. This naturally mitigates the threat of attribute inference. 

Embedding inversion attack directly recovers input text from embeddings and has been shown to recover a certain proportion of input from BERT~\cite{devlin2018bert} embeddings. However, we reveal that embedding inversion attacks achieve low accuracy in recovering prompts from LLM activations. The fundamental challenge comes from \emph{strong non-linearity} of LLMs due to deep depth and large size. We discover in our experiments that the non-linearity of LLM strengthens with the increase of model depth. The strong non-linearity leads to the existence of \emph{numerous input embedding vectors corresponding to similar output activations}, most of which map to meaningless text. The recovered results of embedding inversion attacks on LLMs often fall into these local minima.  
The ineffectiveness of existing attacks towards collaborative LLM inference makes whether these systems are susceptible to privacy threats remains unclear. 

In this paper, to validate the severity of privacy threats in LLM collaborative inference, we define and reveal the prompt inversion attack (\textbf{PIA}),  where a malicious participant intends to recover the input prompt through the activation transmitted by its previous participant. 
In particular, clients mostly request for open source LLMs or parameter-efficiently fine-tuned LLMs when using collaborative inference in practice.  
Accordingly, we consider two representative PIA settings: white-box and grey-box. In the white-box setting, the adversary is assumed to have information about all model parameters; while in the grey-box setting, the adversary only knows the parameters of the base model and his own part of fine-tuned weight.

Under the white-box setting, we design a two-phase attack pipeline, \ie, constrained optimization with adaptive discretization, to resolve the challenge of numerous semantically meaningless local minima due to LLM's strong non-linearity.
Specifically, in the first \emph{constrained optimization} phase, we optimize the input embedding with a constraint term derived from the LLM's embedding matrix. 
This constraint enforces the optimized embedding to be close to the ground truth.
In the second \emph{adaptive discretization} phase, we first construct an optimal attack scheme by adaptively recovering tokens based on activation. However, it inevitably has intractable complexity. We accelerate it by reducing the token candidate set based on the previously optimized embedding. Besides, to enhance the readability of recovered text, we design \emph{semantic speculation} technique based on language prior. 
 For grey-box PIA, due to the lack of knowledge of LoRA parameters, the attacker has to jointly reconstruct the input prompt and estimate the LoRA parameters. We extend our prior white-box PIA method by developing an alternative optimization-based strategy. Our approach iteratively optimizes the generated token and LoRA parameters, making PIA feasible under the grey-box setting.

To reveal the threats of PIA, we conduct extensive experiments on multiple benchmark datasets (\ie, Skytrax~\cite{skytrax}, CMS\cite{cms}, and ECHR~\cite{echr}) and representative LLMs (\ie, Llama-2-70B~\cite{llama2}, Llama-65B~\cite{llama}, and OPT-66B~\cite{opt}). In particular, we have two critical observations: \textbf{(1)} PIA could achieve surprisingly high recovery accuracy on LLMs. 
For example, our method achieves a $88.4\%$ accuracy on the Skytrax dataset and Llama-65B model for the most challenging last participant setting for four participants, whereas the best baseline method adapted from Li et al.~\cite{Generative-embedding-inversion-attack} only achieves $22.8\%$ accuracy. \textbf{(2)} PIA remains highly effective under the grey-box setting. For example, on the Guanaco-65B model fine-tuned on Llama, we achieve a token accuracy of $85.4\%$. These findings demonstrate the vital privacy threat of our attack against the collaborative inference of LLMs.

We also explore whether potential defenses can practically reduce our PIA threats. In general, we find that classical defenses either severely harm generation quality or incur unbearable overhead. For example, our method still retains $79.5\%$ accuracy under the defense of obfuscating activation with Gaussian noises of $\sigma=1$. It further confirms the threats of our PIA attacks.

In conclusion, our main contributions are four-fold. \textbf{(1)} We explore the privacy risks of LLMs under the collaborative inference setting. In particular, we reveal that similar existing techniques (\eg, embedding inversion attacks) are ineffective under this emerging yet practical scenario. 

\textbf{(2)} We formulate and design prompt inversion attack (PIA) to demonstrate the severity of privacy threats in LLM collaborative inference. It is designed under both white-box and grey-box settings and is resistant to potential defenses. 

\textbf{(3)} We provide theoretical insights of our method. Specifically, we present the construction of an theoretically optimal PIA method and further accelerate it to derive our scheme.  We also discover and prove the fundamental weakness of gradient vanishing in previous works [19] as part of our design insights. 

\textbf{(4)} We conduct extensive experiments on benchmark datasets, which verifies the threats of our attack. 
Our white-box attack can achieve nearly perfect recovery of the input prompt, obtaining $88.4\%$ token accuracy under the most challenging case. It significantly outperforms the best baseline (with $22.8\%$ accuracy). We also show that our method can achieve a high $85\%$ token accuracy under the grey-box setting.

\section{Background and Related Work}
\subsection{Large Language Model (LLM)}
The large language model (LLM) is a type of large-scale neural network capable of generating human-like text and completing various tasks. We hereby briefly introduce three common concepts  in LLMs.

\noindent \textbf{Models.} Currently, most advanced LLMs (\eg, GPT-4~\cite{gpt4}) are built upon the transformer architecture~\cite{vaswani2017attention}. These auto-regressive models predict subsequent words in a sequence based on the preceding context. Given a sequence $x_1, x_2, \ldots, x_n$, the model predicts the next word $x_{n+1}$ based on previous words. An auto-regressive LLM consists of one embedding layer, $d$ stacked transformer layers, and one unembedding layer. Each token is represented by a positive integer. The embedding layer is denoted as $E$, mapping $s$-length token sequence to  $\mathbb{R}^{s\times h}$, where $h$ is the hidden state size for each token.  The $i$-th transformer layer is denoted as $T_i$, mapping $\mathbb{R}^{s\times h}$ to $\mathbb{R}^{s\times h}$. The unembedding layer maps the activation of $\mathbb{R}^{s\times h}$ to a single generated token.

\noindent \textbf{Prompt.} A prompt in the context of LLMs refers to the initial model input, guiding the context generation of LLMs. For example, if one inputs a prompt like "Briefly describe how to learn Python", the model will generate a   response telling how to learn python. Prompts are crucial in steering the model's generation process and can vary from simple queries to complex instructions.

\noindent \textbf{LLM Personalization.} There are two major ways for personal users to personalize their own LLMs. The first way is \emph{in-context learning}, in which they generate prompts to describe their personal writing habit, style, and requirement for the task, 
then combine the prompts  with the actual instructions before entering the LLM. The second way is \emph{fine-tuning}. Due to the prohibitive cost of full-weight fine-tuning, personal users normally conduct \emph{parameter-efficient fine-tuning} (only tuning a small proportion of model parameters) on open source LLMs to obtain LLMs with their own style and finish their own task. The most prevalent \emph{parameter-efficient fine-tuning} method is LoRA~\cite{lora}. Under this paradigm, during fine-tuning, only a small proportion of model layers named adapters are updated, while other parameters of the LLM stay frozen.

\subsection{Collaborative Inference System of LLMs}

LLMs pose significant memory and computational costs as they generally have hundreds of billions of parameters. To democratize LLMs, collaborative inference systems~\cite{jiang2023hexgen,PETALS,localai_deml,NESA,skyserve} have been proposed to distribute computation across multiple participants using pipeline parallelism~\cite{pipeline-pl}. 

Recall that an LLM comprises an embedding layer, $d$ stacked transformer layers, and an unembedding layer. The embedding and unembedding layers are lightweight and are thus directly held by the first and the last participant, respectively. Different strategies can be used for mapping transformer layers to participants (nodes), depending on their computing capabilities.  For simplicity, we assume that the transformer layers are evenly distributed among the $n$ participants. Concretely, the first participant holds layers $1, 2, \ldots, \lceil\frac{d}{n}\rceil$.

A significant subset of LLM collaborative inference systems~\cite{edgecloud2,edgecloud3,edgecloud1} involves edge-cloud collaboration~\cite{xu2023survey}. These systems can be considered a specific case of LLM collaborative inference with $n=2$, where the participants are the edge device and the cloud server. Since they face similar risks of prompt leakage, we do not consider them separately in this paper.

Given the prohibitive expense of full-weight fine-tuning LLMs for personal users, we primarily focus on collaborative inference of open-source LLMs or parameter-efficiently fine-tuned open-source LLMs.  Note that for proprietary LLMs like GPT-4~\cite{gpt4}, clients call them through their APIs instead of collaborative inference. 
In the first scenario of open-source LLM inference, we assume that every participant has access to the entire model's weights. In the latter case of parameter-efficiently fine-tuned LLM inference, we assume LoRA~\cite{lora} is adopted due to its prevalence. Under this scenario, before collaborative inference starts, only the client holds his personalized LoRA adapter weights, each participant holds the base model weight which is public. The client distributes the partition of LoRA adapter weights to each participant. Thus, each participant does not know the LoRA weights of other participants.

It is worth noting that collaborative inference is fundamentally different from federated learning~\cite{zhang2021survey}. Firstly, collaborative inference focuses on the inference stage of LLMs, whereas federated learning targets the training phase of models. Secondly, clients (participants) transfer gradients in federated learning, but activation tensors are exchanged in collaborative inference. In addition, each client in federated learning holds the whole model, compared to a small part of the model in collaborative inference. These differences make it impossible for privacy attacks against federated learning to be used directly to attack the collaborative inference of LLMs.

\subsection{Inversion Attacks in Machine Learning}
\label{sec:inversion}
Inversion attack is one of the most critical privacy threats, aiming at recovering input data through information such as intermediate results of a model. In this section, we introduce existing inversion attacks targeting image and language domains.

\noindent \textbf{Inversion Attacks on Images.} Inversion attacks were originally proposed to reconstruct input images. Extensive research has been conducted in this area. In general, existing approaches fall into two categories: generation-based inversion~\cite{yang2019neural,tan2021many,khosravy2022model,yin2023ginver}, and optimization-based inversion~\cite{mahendran2015understanding,he2019model,attacking-and-protect-data-privacy,zhang2020secret,struppek2022plug}. 
\cite{he2019model} studied inverting CNN intermediate activations by optimizing a loss term combining total variation to encourage the inverted image to look realistic. 
\cite{zhang2020secret} proposed to optimize GAN's latent vector space to produce high-quality inversion results. 
However, these methods are not applicable on texts because images can be represented in a continuous space while words in language modality is discrete. This makes the inversion more challenging for language compared with image.

\noindent \textbf{Inversion Attacks on Language.}
Most inversion attacks on language models focused on recovering sensitive information from word or sentence embeddings.  \cite{song2020information} proposed formulating language embedding inversion as an optimization problem. Several works~\cite{Sentence_inference_attack, Generative-embedding-inversion-attack} trained an extra decoder model to invert sentence embeddings by generation. \cite{pan2020privacy} focused on recovering partial information. 

However, as we will show in later sections, a substantial reason for the ineffectiveness of these methods in recovering a long prompt from LLM activations lies in the strong non-linearity of LLMs. Moreover, \cite{song2020information} has limited performance on LLMs because of its simple discretization procedure which lacks the guidance of ground-truth activation. 
Generation-based methods~\cite{Generative-embedding-inversion-attack,Sentence_inference_attack} only incorporate activation information at the start of auto-regressive decoding. When the generation length increases, it no longer recovers sentences precisely because the dependence between recovered tokens and the activations gradually disappears. 



Recent work~\cite{morris2023language} considered another setting of recovering prompts from LLM output logits (output token probability). 
Although their method could be applied under our white-box setting, its  recover accuracy is also very limited. We speculate that the poor performance is because their inversion is also based on a trained  decoder model. Similar to the drawbacks of generation-based embedding inversion methods discussed above,  their recovery accuracy drastically decreases when recovering long text.  

As such, whether LLM collaborative inference is susceptible to inversion attacks remains an open question.




\section{Threat Model and Problem Formulation}
In this section, we illustrate the threat model of our prompt inversion attack (PIA) against LLM collaborative inference systems and delve into formulations of PIA.

\subsection{Threat Model}
\label{sec:threat_model}
This paper focuses on the privacy of clients' prompts because they often carry sensitive information. Without loss of generality, we focus on the case of one malicious participant. The attack is easier to conduct if multiple participants are compromised by a single attacker due to the acquisition of more information. Hence, we consider this to be the most challenging case.  We omit the case that the first participant is malicious because the first participant is allocated to be the client himself. This eliminates the need for the client to send the prompt to the first participant, thereby preventing direct prompt leakage.  Following the concept in secure multi-party computation (MPC)~\cite{yao1986generate,beerliova2008perfectly}, we model the attacker as honest-but-curious, \ie, he faithfully computes his layers and sends the correct activation to the next participant, as explained in Figure~\ref{fig:concept}.

\revise{Since proprietary LLMs are unsuitable for collaborative inference, which requires parties to hold parts of the model weights, in our paper, we mainly focus on the setting of collaborative inference of a publicly known model (\emph{white-box setting}).  Arguably, white-box setting dominates the real-world applications of collaborative inference because open-source LLMs like LlaMA-3 and DeepSeek~\cite{deepseek} are becoming increasingly powerful, achieving performance comparable to proprietary LLMs.}
Besides, we also explore the attack of models fine-tuned with parameter-efficient fine-tuning (PEFT) techniques (\emph{grey-box setting}). Without loss of generality, we take the most prevalent PEFT technique LoRA~\cite{lora} as an example in our paper.

LoRA~\cite{lora} hypothesizes that during fine-tuning, updates to the weights of the base model exhibit a low ``intrinsic rank".   For a pre-trained weight matrix $W_i \in \mathbb{R}^{d \times k}$ at the $i$-th transformer layer, the weight update is constrained by a low-rank decomposition  $W_i+\Delta W_i=W_i+\theta_{B_i} \theta_{A_i}$, where $\theta_{B_i} \in \mathbb{R}^{d \times r}, \theta_{A_i} \in \mathbb{R}^{r \times k}$, and the rank $r \ll \min (d, k)$. During fine-tuning, $W_i$ remains frozen without receiving gradient updates, while $\theta_{A_i}$ and $\theta_{B_i}$ contain trainable parameters. In the collaborative inference for LoRA fine-tuned LLMs, each participant  receives their respective  $\theta_{A_i}, \theta_{B_i}$ from the client and computes their updated weight matrix $W_i+\theta_{B_i} \theta_{A_i}$. 

For the white-box setting, the attacker knows the whole model's weights. 
For the grey-box setting, all participants know the base model's weights and  their own LoRA weight, but have no information about the LoRA weights of others.

We assume the attacker records the received activation during collaborative inference silently. After recording, the attacker conducts the inversion. Note that the inversion can be done completely offline and has no urgent timeliness requirements. Thus we assume  the attacker has enough computational power to perform the inversion. Performing the attack can be profitable due to two reasons: (1) Input prompts
often contain sensitive information such as personal names and
contact information. (2) High quality LLM prompts themselves
 have commercial value and can be traded on markets~\cite{promptbase,promptmarket}.


\subsection{Problem Formulation} 
Assume the attacker is the $i$-th participant. Under the \emph{white-box setting}, the attack problem can be formulated as follows. The attacker receives the previous activation $A$ and he knows the model layer parameters of all preceding  participants'  $F_1, F_2, \ldots, F_{i-1}$.  The relationship between $A$ and $x$ is:
\begin{align}
    A= F_{i-1}\circ F_{i-2} \circ \cdots \circ F_{1} (E(x)).
\end{align}
where $E(\cdot)$ is the embedding layer. The attacker's objective can be formulated as: for each input prompt $x$, given $A, F_1, F_2, \ldots, F_{i-1}$, find $x'$, such that $x'\approx x$.

Under the \emph{grey-box setting}, the attack problem can be formulated as follows. We define $G(\cdot , \cdot)$ as a model layer fine-tuned with LoRA, which includes the layer's base weights $w_i$ and its LoRA adapter $\theta_{A_i},\theta_{B_i}$. The attacker receives the previous activation $A$ and he has access to all previous participants' model base weights $w_1, w_2, \ldots, w_{i-1}$. His goal is to recover the input prompt $x$. The relationship between $A$ and $x$ is:
\begin{footnotesize}
        \begin{align}
    A= G_{w_{i-1},\theta_{A_{i-1}},\theta_{B_{i-1}}}\circ \cdots \circ G_{w_1,\theta_{A_1},\theta_{B_1}} (E(x)).
\end{align}
\end{footnotesize}

The attacker's objective can be formulated as: for each input prompt $x$, given $A, G, w_1, w_2, \ldots, w_{i-1}$, find $x'$, such that $x'\approx x$.
\section{Observations and Insights}
\label{sec:revisit}

Given the limitations of  generation-based attacks, we favor optimization-based ones for our design. 
In this section, we revisit two representative optimization-based attacks: He et al.~\cite{he2019model} in the image domain and Song et al.~\cite{song2020information} in the language domain. We have the following two findings:  

\noindent
\textbf{(1)} 
\cite{he2019model} inspires us to exploit prior knowledge on input distribution. It improves inversion by increasing the probability of the optimized result being close to the ground truth.

\noindent
\textbf{(2)} A major reason why~\cite{song2020information} does not perform well on LLM activation inversion is gradient vanishing. We theoretically prove that the vanishing phenomenon is intrinsic due to the adoption of softmax attention mechanism in its optimization. 






We firstly revisit an optimization-based inversion attack on image domain~\cite{he2019model}. Denote the model layers as $f$ and the ground-truth input as $x_0$. Given the activation $f(x_0)$, attackers intend to find an image sample $x^*$ satisfying two requirements: \textbf{(1)} its activation $f(x^*)$ is similar with $f(x_0)$; \textbf{(2)} $x^*$ follows the distribution of natural images.

For the first requirement, \cite{he2019model} used the Euclidean distance to measure the similarity between $f(x)$ and $f(x_0)$. For the second requirement, \cite{he2019model} exploited the total variation (TV)~\cite{rudin1992nonlinear} to represent natural image prior. 
Minimizing TV encourages  $x$ to be piece-wise smooth, therefore looking realistic. \cite{he2019model} designed their optimization function by linearly combining the two goals of maximizing similarity between $f(x),f(x_0)$ and minimizing $TV(x)$. 

The goal of minimizing TV is crucial for image recovery in~\cite{he2019model} because, non-linearity of neural networks may result in many local minima of $\|f(x)-f(x_0)\|_2$, which satisfy $f(x)\approx f(x_0)$. However, most local minima are garbled images far from $x_0$. Without this constraint on TV, 
starting optimization with randomly initialized points is very likely to obtain $x$ that corresponds to garbled images.  Adding this constraint mitigates the non-linearity problem by limiting the search space of $x$ within low TV (look realistic). The TV constraint filters out the most of these meaningless local minima, thereby increasing the likelihood that  $x$ will be close to $x_0$.

We can learn from~\cite{he2019model} that \emph{enforcing the prior on input data distribution is critical for improving inversion}. This is because the input prior can increase the likelihood of the inversion result being close to the ground-truth input.

Secondly, we revisit~\cite{song2020information}, the only recent work studying optimization-based inversion attacks for language models, to the best of our knowledge. While their techniques appear compatible with our setting, substantial modifications are required for accuracy. 

Denote the language model transformer layers as $\Phi(\cdot)$, embedding layer as $E$. \cite{song2020information} formulated the white-box inversion as the following optimization problem:
\begin{align}
\label{eq:song_formulate}
\min _{\hat{x} \in \mathcal{X}(\mathcal{V})}\left\|\Phi(E(\hat{x}))-\Phi(E(x))\right\|_2^2,
\end{align}
 where $x$ is the ground-truth input embedding, $\Phi(E(x))$ represents the target activation, $\hat{x}$ denotes inverted text, and $\mathcal{X}(\mathcal{V})$ denotes all possible sequences using vocabulary $\mathcal{V}$.

To bridge the gap between discrete space and continuous space, they relax the word selection at each position  with a continuous variable $z_i \in \mathbb{R}^{|\mathcal{V}|}$.  Input embedding  $\hat{v}_i$ is computed by $z_i$ using the softmax function on the embedding matrix $V$:
\begin{align}
\label{eq:softmax_attn}
\hat{\boldsymbol{v}}_i=V^{\top} \cdot \operatorname{softmax}\left(z_i / T\right), \text { for } i=1, \ldots, \ell,
\end{align}
where $T$ is a temperature parameter. The softmax function approximates hard argmax selection for $T\le1$. Let $z=\left[z_1, \ldots, z_{\ell}\right] \in \mathbb{R}^{\ell \times|\mathcal{V}|}$. The relaxed problem is:
\begin{align}
\label{eq:relaxed_problem}
\min _z\left\|\Phi([\hat{\boldsymbol{v}}_1, \ldots, \hat{\boldsymbol{v}}_{\ell}])-\Phi(x)\right\|_2^2.
\end{align}

With white-box access to $\Phi$, the optimization problem is solved by gradient descent. After optimization, the optimized embedding $\hat{\boldsymbol{v}}$ and corresponding score vector $z$ are obtained. For text, it is necessary to discretize the optimized embedding back into token space. To recover the discrete tokens from embedding space, we compute $\hat{x}=\left\{w_i \mid i=\arg \max z_j\right\}_{i=1}^{\ell}$, selecting each position's word based on the similarity in the input embedding space.

However, we have empirically found that \emph{its performance significantly decreases with large models}, rendering it ineffective for solving our problem.  
For instance, we evaluate it on inverting a Llama-7B model. Results in Table~\ref{tab:ccs20_Llama_7b} show that \cite{song2020information} can barely recover tokens based on the activations of the Llama model.

\begin{figure}[!t]
\centering
\subfloat[$|\frac{\partial L}{\partial z}|$ \label{fig:zgrad}]{\includegraphics[width=0.25\textwidth]{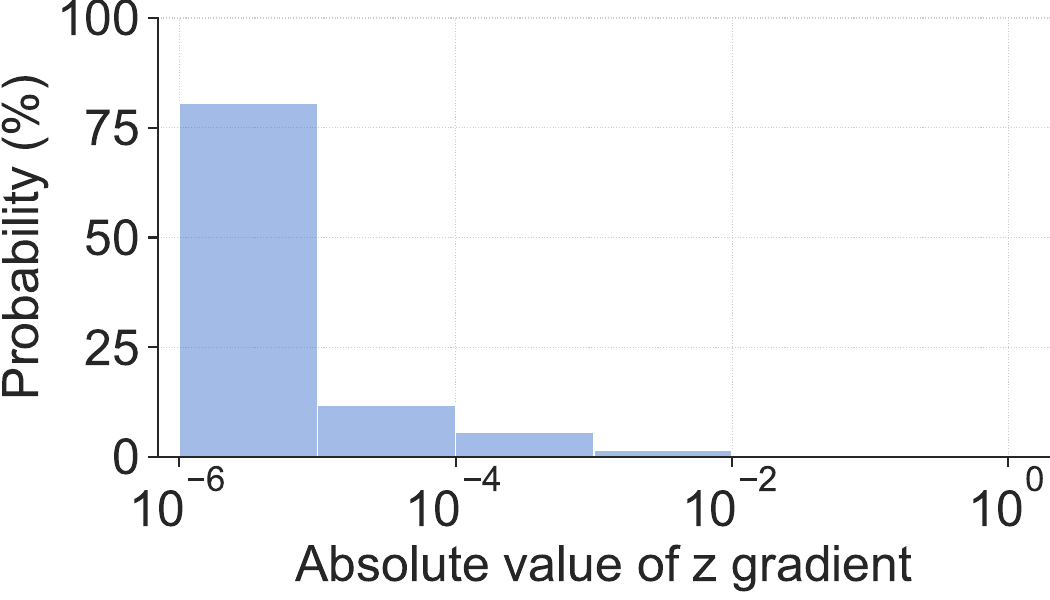}}
\subfloat[$|\frac{\partial L}{\partial \hat{v}}|$ \label{fig:vgrad}]{\includegraphics[width=0.25\textwidth]{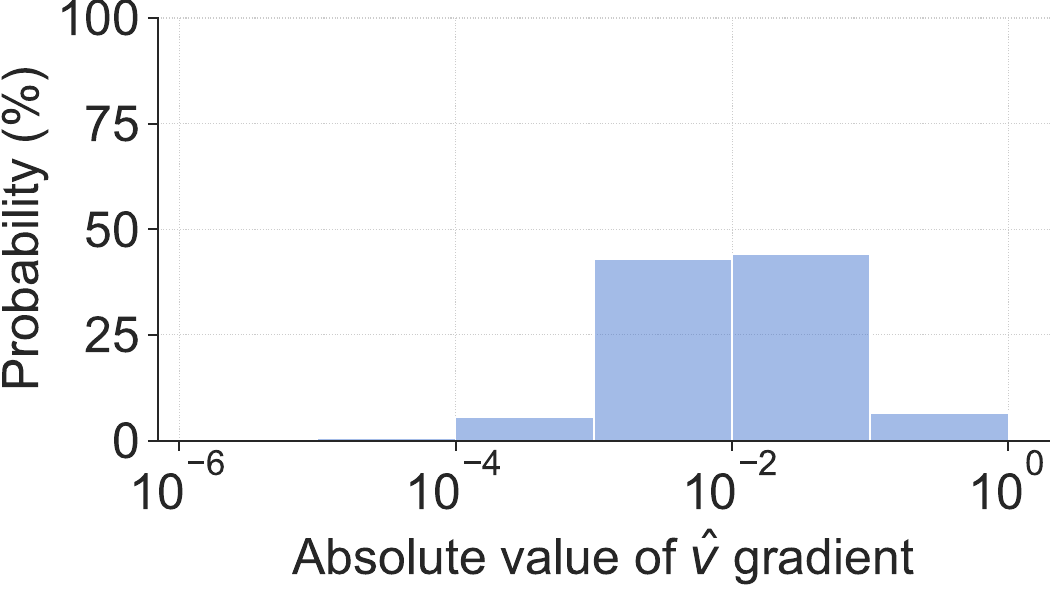}}
 \vspace{-0.5em}
\caption{The distribution of $|\frac{\partial L}{\partial z}|$ and $|\frac{\partial L}{\partial \hat{v}}|$.}
\label{fig:z_vhat_grad}
 \vspace{-1em}
\end{figure}

We further investigate the reason of \cite{song2020information}'s  limited performance and discover that \emph{gradient vanishing caused by the softmax attention mechanism is a crucial reason}. Denote the optimization goal in Equation (\ref{eq:relaxed_problem}) as $L$, we visualize the value distribution of $\frac{\partial L}{\partial \hat{v}}$ and $\frac{\partial L}{\partial z}$ for Llama-7B model in Figure~\ref{fig:z_vhat_grad}.  Figure~\ref{fig:zgrad} shows that $z$'s gradient vanishes, and most gradient values are almost zero. In contrast, Figure~\ref{fig:vgrad} shows that the gradient of $\hat{v}$ does not vanish. Recall that $\hat{v}$ is computed by $z$ with  softmax in Equation (\ref{eq:softmax_attn}), it can be inferred that the softmax attention mechanism causes gradient vanishing. 

In particular, we prove that \emph{the previous vanishing phenomenon is intrinsic} rather than accidental, as follows.


\begin{theorem}
\label{thm:vanish}
For vocabulary set $V$, embedding hidden dimension $h$, token embedding matrix $W:\mathbb{R}^{|V|\times h}$, embedding vector $v:  \mathbb{R}^{h}$ , word selection weight $z:\mathbb{R}^{|V|}$, optimization target value $L$, $v=W^\top softmax(z)$. Let $s=softmax(z)$. Assume $W_{ij} \sim \mathcal {N}(0,\sigma^2)$ for $i\in \{1,2,\cdots,|V|]\}, j \in \{1,2,\cdots,h\}$. With high probability, the following inequality holds:\\
\begin{align}
    ||\frac{\partial L}{\partial z} ||_1\le 4  \sigma \sqrt{\sum_{j=1}^h (\frac{\partial L}{\partial v_j})^2}
\end{align}

\end{theorem}

In general, Theorem~\ref{thm:vanish} indicates that the $L_1$ norm of the gradient is upper-bounded by a small value ($\sigma$ is small in practice) with high probability, which proves the vanishing phenomenon. 
The proof of Theorem \ref{thm:vanish} and more numerical analysis can be found in Appendix~\ref{appendix:ccs20_vanish_theory}.

%



Drawing from these observations, we gain critical insights that lead us to impose constraints on input during optimization attacks in the text domain, eliminating the softmax attention mechanism in~\cite{song2020information}. These strategic adjustments effectively tackle the challenges identified in~\cite{song2020information}.  Technical details of our method are presented in the following section. 



\begin{table}[t]
\centering
\scalebox{1.4}{
\begin{tabular}{|c|c|c|c|}
\hline
Layer number   & 4   & 8 & 16 \\ \hline
Token accuracy(\%) & 0.4 & 0 & 0  \\ \hline
\end{tabular}
}
\caption{Token accuracy of \cite{song2020information} for inverting Llama-7B on Skytrax~\cite{skytrax} dataset.}
\label{tab:ccs20_Llama_7b}
 \vspace{-1em}
\end{table}

\section{Methodology}

\label{sec:method}

\begin{figure*}
\centering
\includegraphics[width=0.8\textwidth]{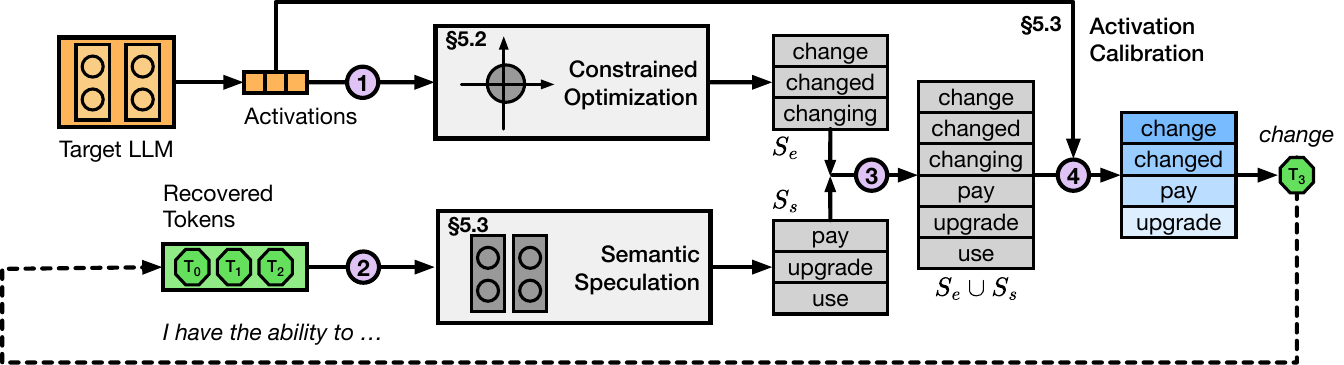}
 \vspace{-0.5em}
\caption{Workflow of our prompt inversion attack. \textbf{(1)} The attacker first conducts \textit{constrained optimization} to approximately recover the  input embedding, and obtain the  embedding-distance based candidate set $S_e$. Our designed constraint improves the result quality of  embedding optimization. 
The attacker also computes a semantic-based candidate set $S_s$, obtained by feeding previously recovered tokens into an oracle LLM. \textbf{(2)} $S_e$ and $S_s$ are then joined. \textbf{(3)} Activation calibration. \textbf{(4)}  The attacker enumerates the tokens in $S_e \bigcup S_s$ and finds the token that best matches the target activation.
}
\label{fig:pipeline}
 \vspace{-1em}
\end{figure*}

\begin{figure}[!t]
\centering
\includegraphics[width=0.4\textwidth]{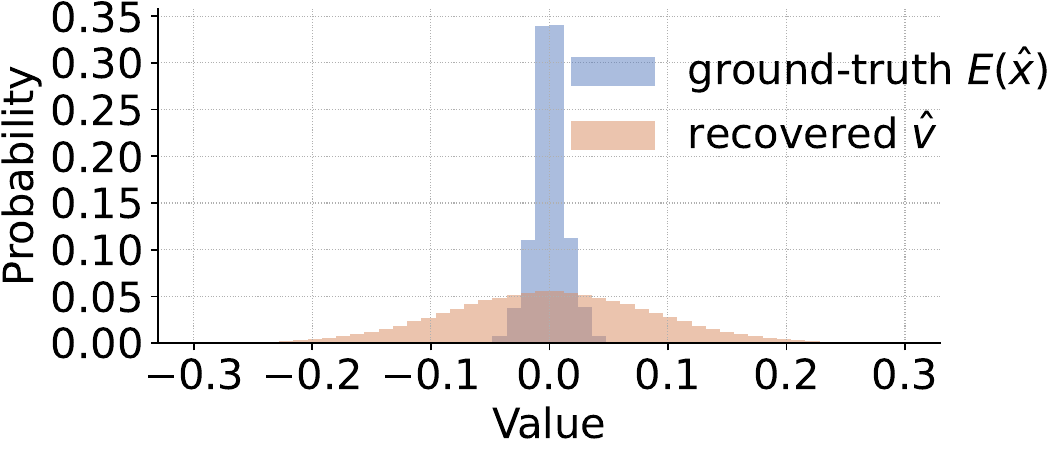}
 \vspace{-0.5em}
\caption{Value distribution of ground-truth embedding $E(x)$ and  optimized embedding $\hat{v}$ obtained by solely optimizing based on Equation (\ref{eq:easier version}) on Llama-7B model. 
  We can observe that $\hat{v}$ is far away from $E(x)$. } 
 \vspace{-1em}
\label{fig:value_distribution}
\end{figure}

\begin{figure*}
\centering
\vspace{-1em}
\includegraphics[width=0.9\textwidth]{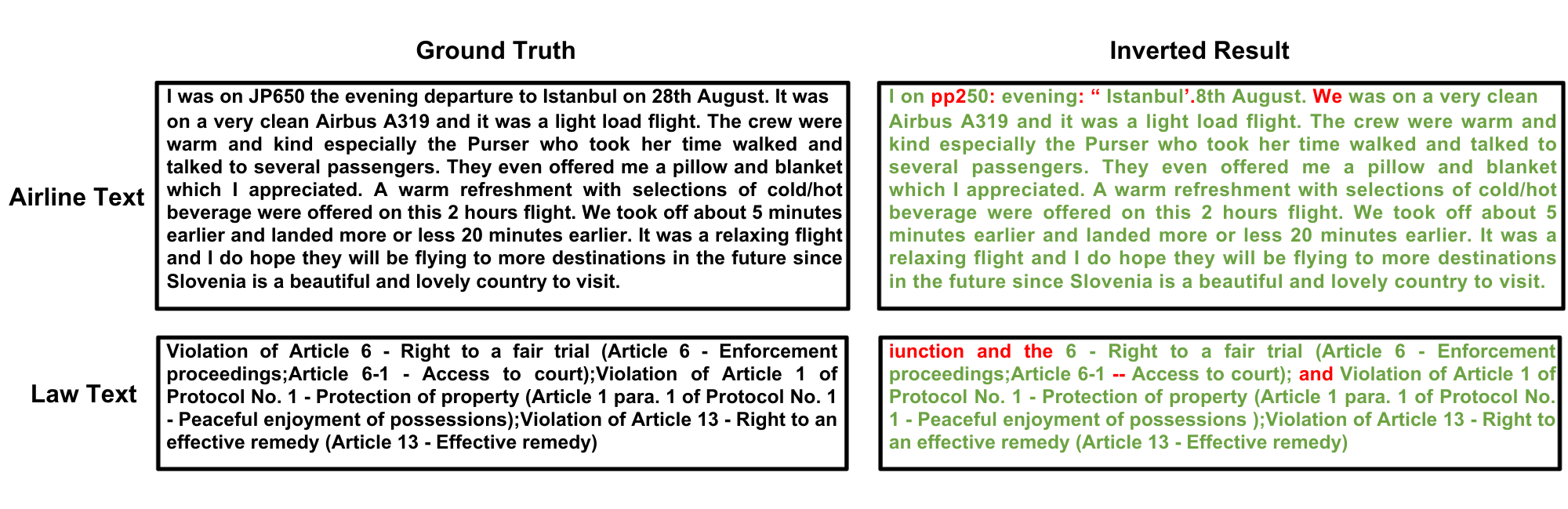}
\caption{Two examples of our prompt inversion attack. The first example prompt is an airline review from Skytrax dataset, while the second is from ECHR Law dataset. Tokens inverted correctly are marked in green, while the wrong ones are marked in red. } 
\label{fig:example}
 \vspace{-1em}
\end{figure*}

In this section, we elaborate on how we derive and design an effective prompt inversion attack. 

\subsection{Outline of Prompt Inversion Attack}
In this section, inspired by discoveries in Section~\ref{sec:revisit}, we design a new optimization-based prompt inversion attack. In general, our method consists of two phases: \textit{constrained optimization} and \textit{adaptive discretization}, as shown in Figure~\ref{fig:pipeline}. The whole pipeline of our white-box PIA is elaborated in Algorithm~\ref{alg:core} in Appendix~\ref{appendix:alg}. 

Specifically, we formulate prompt inversion attack as the following problem: find $\hat{x}\in \mathcal{X}(\mathcal{V})$,  s.t.,
\begin{align}
\label{eq:first problem}
\begin{aligned}
     \min  \| F(E(\hat{x}))-F(E(x)) \|_2^2. \\ 
\end{aligned}
\end{align}
where $x$ is the ground-truth token sequence, $F$ is the previous model layers, $E$ is the embedding layer, and $\mathcal{X}(\mathcal{V})$ denotes all possible sequences using vocabulary $\mathcal{V}$.  

Considering the difficulty of discrete optimization, we relax the problem above by adding a continuous variable representing the input embedding. The problem is transformed to: find $\hat{v} \in \mathbb{R}^{|x|\times h}, \hat{x}\in \mathcal{X}(\mathcal{V})$,  s.t.,
\begin{align}
    & \min_{\hat{v}}  \| F(\hat{v})-F(E(x)) \|_2^2, \label{eq:easier version} \\ 
    & \min_{\hat{x}}  \| E(\hat{x})-\hat{v} \|_2^2  .\label{eq:easier version_constraint}
\end{align}
To avoid gradient vanishing, our method directly optimizes on the input embedding without the softmax mechanism in \cite{song2020information}. 
A straightforward solution for the above problem is  first leveraging Equation (\ref{eq:easier version})  to optimize $\hat{v}$ then discretizing $\hat{x}$ by Equation (\ref{eq:easier version_constraint}). 

However, this solution leads to poor accuracy. 
This is due to the strong non-linearity of LLMs:  Equation (\ref{eq:easier version}) exhibits numerous local minima, many of which are far from the ground-truth. Therefore, directly optimizing Equation (\ref{eq:easier version}) from random starting points is highly likely to converge to these  local minima distant to  ground-truth.

To address this problem, we propose \textit{constrained optimization}. Specifically, \textit{constrained optimization} constrains the search space of $\hat{v}$ by introducing a constraint term derived from the embedding matrix. Combining the constraint in optimization filters out most local minima of Equation (\ref{eq:easier version}) that are far away from $E(x)$. Thus the constraint significantly increases the probability of the optimized embedding $\hat{v}$ to be close to ground-truth.

After recovering the input embedding by \textit{constrained optimization}, we aim to push our token recovery accuracy to the extreme. To achieve this, we propose \textit{adaptive discretization}. This method selects a small token candidate set based on optimized embeddings, then adaptively discretizes the optimized embedding based on ground-truth activation information.
Furthermore, to enhance the readability of recovered text, we propose 
 \textit{semantic speculation} by integrating natural language semantic coherence prior. 





\subsection{Constrained Optimization}

As illustrated previously, a straightforward way to design a PIA is to solve Equation (\ref{eq:easier version}) and Equation (\ref{eq:easier version_constraint}) independently and separately. Specifically, we can first independently optimize Equation (\ref{eq:easier version}) by gradient descent to obtain $\hat{v}$ (named as \textit{naive optimization}); then solve Equation (\ref{eq:easier version_constraint}) to obtain $\hat{x}$ by finding each $\hat{v}_i$'s nearest token in embedding space, named as \textit{naive discretization}. 

However, as shown in Figure~\ref{fig:value_distribution}, $\hat{v}$ obtained by naive optimization is (numerically) far away from the ground-truth embedding $E(x)$. This is rooted in the strong non-linearity of $F$, which leads to the existence of numerous local minima of $\hat{v}$ that are mostly distant from the ground-truth embedding $E(x)$.  Optimizing Equation (\ref{eq:easier version}) from random starting  points is highly likely to fall into these local minima, ultimately resulting in poor inversion accuracy. 

To reduce the distance between optimized embedding $\hat{v}$ and ground-truth, a straightforward idea is to constrain this distance during optimization.  Based on this idea, we consider jointly optimizing  Equation (\ref{eq:easier version}) and  $\|\hat{v}-E(x)\|_2$.




\noindent \textbf{Joint Optimization.} To jointly consider the two goals, we apply Lagrange multiplier method, as follows:
\begin{align}
\label{eq:lagrange version}
\begin{aligned}
    & \min_{\hat{v}}  \| F(\hat{v})-F(E(x)) \|_2^2+ \lambda \|\hat{v}-E(x)\|_2^2,
\end{aligned}
\end{align}
where $\lambda$ is the Lagrange multiplier. 





Unfortunately,  Equation (\ref{eq:lagrange version}) is not directly solvable by the attacker, because he only knows $F(E(x))$ while has no access to $E(x)$. This makes computing the second term in Equation (\ref{eq:lagrange version}) infeasible. To address the problem,  we relax the second term in Equation (\ref{eq:lagrange version}) with its lower bound. Firstly, we have the following relation: 
\begin{align}
\label{eq:lagrange each term}
\begin{aligned}
    &  \| \hat{v}_i- E(x_i) \|_2^2\ge \min_{t \in \mathcal{V}} \|\hat{v}_i -E(t) \|_2^2.
\end{aligned}
\end{align}
This relation lower bounds the distance between $\hat{v}_i$ and $E(x_i)$ by enumerating over all possible tokens $t$.

Furthermore, we can lower bound the second term in Equation (\ref{eq:lagrange version}) by applying Inequality~(\ref{eq:lagrange each term}):\\
\begin{align}
\label{eq:lower bound}
\begin{aligned}
    &  \| \hat{v}- E(x) \|_2^2\ge \sum_{i=1}^{|x|} \min_{t \in \mathcal{V}} \|\hat{v}_i -E(t) \|_2^2.
\end{aligned}
\end{align}

\noindent \textbf{Overall Optimization Objective.} Substituting by the term in Inequality (\ref{eq:lower bound}), the optimization goal in Equation (\ref{eq:lagrange version}) could be relaxed to :
\begin{align}
\label{eq:whole lagrange practical version}
\begin{aligned}
    \min_{\hat{v}}  \| F(\hat{v})-F(E(x)) \|_2^2+ \lambda \cdot \sum_{i=1}^{|x|} \min_{t \in \mathcal{V}} \|\hat{v}_i -E(t) \|_2^2.
\end{aligned}
\end{align}

The intuition here is that, we relax the second target from constraining $\hat{v}$ to be close to $E(x)$ (which is infeasible to solve since the attacker does not know $E(x)$), into constraining $\hat{v}$ to be close to its nearest legitimate token sequence's embedding.

We can obtain $\hat{v}$ by optimizing Equation (\ref{eq:whole lagrange practical version}) with gradient descent. Another trick to further constrain $\hat{v}_{ij}$ is to  clip all $\hat{v}_{ij}$ with $L_j$ and $R_j$  at each optimization step, where $L_j=\min_{t \in \mathcal{V}} E(t)_j, R_j=\max_{t \in \mathcal{V}} E(t)_j$. $L,R$ are the lower bounds and upper bounds on each embedding dimension computed from the embedding matrix, respectively. We name this new optimization procedure as \textit{constrained optimization}, whose details are depicted in Algorithm~\ref{alg:optim} in Appendix~\ref{appendix:alg}.



Compared with naive optimization, \textit{constrained optimization} improves by adding an regularization term to enforce the constraint that $\hat{v}$ should be close to its nearest legitimate token sequence's embedding. 
The benefit of this constraint is as follows: If $\hat{v}$ is close to  a token sequence $X$'s embedding (enforces by our constraint), then with high probability $F(\hat{v})$ is close to $F(E(X))$.  The first goal of our optimization leads to $F(\hat{v})$ close to $F(E(x))$. Thus $F(E(X))$ will be close to $F(E(x))$. This is only possible when $X$ is semantically close to $x$. Therefore $\hat{v}$ is likely to be close to $E(x)$ after optimization. 
This explains why our constrained optimization works well in practice. 


\subsection{Adaptive Discretization}
After $\hat{v}$ is obtained by the previous \textit{constrained optimization}, the next step is to recover discrete tokens $\hat{x}$. 
A simple way is to map each $\hat{v}_i$ to its nearest embedding in the vocabulary: $\hat{x_i}=\arg\min_{\hat{x_i} \in \mathcal{V}} \| E(\hat{x}_i)-\hat{v}_i \|_2^2$, named as \textit{naive discretization}.

This method is straightforward, however, it does not seem to obtain the optimal $\hat{x}$. 
The reason is that \textit{naive discretization} only aims to minimize the distance between discretized tokens $\hat{x}$ and the optimized embedding $\hat{v}$ ($| E(\hat{x}) - \hat{v} |_2$). However, our eventual goal is to minimize the distance in the activation space ($| F(E(\hat{x})) - A |_2$), where $A=F(E(x))$.
There is still some gap between these two goals, due to the non-linearity of $F$ and $F(\hat{v})\neq A$.





To achieve the goal of minimizing the distance in activation space, we construct a discretization scheme based on the idea of greedily calibrating with the ground-truth activation information at each position. The construction is as follows:

\noindent \textbf{ Activation Calibration.}  At the $j$-th step, we choose the token from the vocabulary which best calibrates the ground-truth activation at that position. Namely, we recover token $\hat{x}_j$ by choosing the token that minimizes $\| F(E(\hat{x}_1,\hat{x}_2,\cdots,\hat{x}_{j-1},\hat{x}_j))_j-A_j) \|_2^2$.

For this discretization procedure, we discover that it has the intriguing property of \emph{perfectly recovering the input prompt}.   To analyze this property, we firstly derive Theorem \ref{thm:accumulate}.


\begin{theorem}
\label{thm:accumulate}
For any position $j$, let  token $\hat{x}_j$ be computed by:
\begin{align}
\label{eq:sj_def}
\hat{x}_j=\argmin_{\hat{x}_j\in \mathcal{V}}\| F(E(\hat{x}_1,\hat{x}_2,\cdots,\hat{x}_{j-1},\hat{x}_j))_j-A_j \|_2^2 .
\end{align}
$\forall t \in \mathcal{V}$, we have:
\begin{align}
    \begin{aligned}\label{eq:thm2_proof_target}
       &  \| F(E(\hat{x}_1,\hat{x}_2,\cdots,\hat{x}_{j-1},t))-[A_1,\cdots,A_j]) \|_2^2 \\
      \ge  &  \| F(E(\hat{x}_1,\hat{x}_2,\cdots,\hat{x}_{j-1},\hat{x}_j))-[A_1,\cdots,A_j]) \|_2^2 .\\
\end{aligned}
\end{align}

\end{theorem}

We provide the complete proof in Appendix~\ref{appendix:error_accumulate}. The core idea of the proof is to leverage the masked attention mechanism of LLM (\ie, $A_j$ is only determined by $[x_1,\cdots,x_j]$, and is irrelevant with subsequent tokens).

From Theorem~\ref{thm:accumulate}, we can prove that $\hat{x}_j$ is the token  appended to $[\hat{x}_1,\cdots,\hat{x}_{j-1}]$ resulting in an activation nearest to the ground-truth activation $[A_1,\cdots,A_j]$.  Furthermore, at the first step of discretization, $\hat{x}_1$ will be chosen as $x_1$. Then, consecutively, $\hat{x}_2$ is chosen as $x_2$, and so on. Therefore, it can perfectly recover each token of $x$.

\noindent \textbf{Acceleration of Optimal Scheme.} However, the previous scheme is intractable in practice because the computation of each $\hat{x}_j$ incurs running an LLM $\mathcal{O}(|\mathcal{V}|)$ times (enumerating the vocabulary). Thus, its complexity is $\mathcal{O}(|\mathcal{V}|\cdot |x| \cdot C)$, where $C$ is the complexity of running the LLM. 

To accelerate this optimal method, we need to avoid enumerating $\hat{x}_j$ over vocabulary $\mathcal{V}$. After  our \textit{constrained optimization} procedure, $\hat{v}_j$ is likely to be near $E(x)_j$. Based on this intuition, instead of thoroughly enumerating $\mathcal{V}$, we can approximate the perfect discretization by only searching $\hat{x}_j$ in the  $\hat{v}_j$'s top-$K$ nearest tokens. We denote the set of these tokens as \textit{embedding based candidate set} $S_e$. We efficiently invert $\hat{x}_j$ by searching in $S_e$:
\begin{align}
\label{eq:discretize_practical}
    \hat{x}_j=\argmin_{t\in S_e}\| F(E(\hat{x}_1,\hat{x}_2,\cdots,\hat{x}_{j-1},t))_j-A_j \|_2^2 .
\end{align}

\noindent \textbf{Improving Readability via Semantic Speculation.} 
Despite significant accuracy improvement, the above approach still has the drawback of including a certain number of strange tokens and syntax errors in the recovered text. This harms the readability of the recovered text.
The above method is unable to recover $\hat{x}_j$ correctly when $\hat{x}_j$ does not belong in $S_e$, which often happens when $\hat{x}_j$'s meaning is obscure. For example, for proposition words like ``of"  or punctuation, since their semantic meanings are not strong enough, they are easily mixed with other words in embedding space, making the recovery of these tokens difficult. 
Fortunately, leveraging the semantic coherence of natural language, these semantically obscure tokens can be accurately recovered.  We integrate semantic coherence  into discretization as follows: we leverage a standalone LLM, such as Llama-7B~\cite{llama}, to predict the next token given input $[\hat{x}_1,\hat{x}_2,\cdots,\hat{x}_{j-1}]$. Then, we take the top-$Y$ tokens predicted to form the \emph{semantic-based candidate set} $S_s$. We merge $S_e$ and $S_s$ as the total candidate set and enumerate the tokens in  $S_e\bigcup S_s$ to find $\hat{x}_j$. Formally, we invert  $\hat{x}_j$ by:
\begin{align}
    \hat{x}_j=\argmin_{t\in S_e\bigcup S_s}\| F(E(\hat{x}_1,\hat{x}_2,\cdots,\hat{x}_{j-1},t))_j-A_j \|_2^2 .
\end{align}
This technique is named \textit{semantic speculation}. By this approach, when $\hat{x}_j$ fails to fall into $S_e$ but falls into $S_s$, we can still recover $\hat{x}_j$ by activation calibration.

\noindent \textbf{Overall Discretization.} We name the above whole discretization procedure as \textit{adaptive discretization} since we adaptively leverage the ground-truth activation $A$'s information and the semantic coherence prior in the discretization procedure. Details of the \textit{adaptive discretization} procedure are depicted in Algorithm~\ref{alg:discrete} in Appendix~\ref{appendix:alg}. 



\subsection{Grey-box Prompt Inversion Attack}
In this section, we explore how to design an effective grey-box PIA. As described in our threat model in Section~\ref{sec:threat_model}, the attacker does not know the LoRA adapter weights of preceding layers and only knows the base weight.

This problem can be modeled as follows: given the activation $A=G_{w_{a-1},\theta_{A_{a-1}},\theta_{B_{a-1}}}\circ G_{w_{a-2},\theta_{A_{a-2}},\theta_{B_{a-2}}} \circ \cdots \circ G_{w_1,\theta_{A_1},\theta_{B_1}} (E(x))$ and base weight $w_1,w_2,\cdots,w_{a-1}$, find $\hat{x}=\arg\max_{\hat{x}} \|\hat{x}-x\|_0$. The overall formulation can be denoted by:
Find $\hat{x},\theta_{A_1},\theta_{B_1},\cdots,\theta_{A_{a-1}},\theta_{B_{a-1}}$, s.t.,
\begin{small}
    \begin{align}
\label{eq:grey_box_problem}
\hspace{-1em}
    \min \|G_{w_{a-1},\theta_{A_{a-1}},\theta_{B_{a-1}}} \circ \cdots \circ G_{w_1,\theta_{A_{1}},\theta_{B_{1}}} (E(\hat{x}))-A \|_2,
\end{align}
\end{small}
where the attacker simultaneously estimates the LoRA weights  $\theta$ and recovers the private input prompt $\hat{x}$.

The main difficulty of grey-box PIA compared with the white-box one is that grey-box PIA involves multiple unknown vector variables, while in the white-box problem only $\hat{x}$ is unknown. 

To address this challenge, we propose to optimize LoRA weights and input embedding in an alternative optimization manner. Specifically, in each iteration, we firstly fix $\theta$ and only optimize $x$; in this case, the problem is simplified as a white-box PIA, and can be solved using our previous method; then we fix $x$ and optimize $\theta$, which is done by gradient descent: the goal of optimization on $\theta$ is also minimizing activation distance. Specifically, we initialize the LoRA adapters $\theta_{A_1}^{(1)},\theta_{B_1}^{(1)},\cdots,\theta_{A_{a-1}}^{(1)},\theta_{B_{a-1}}^{(1)}$ with zeros. We firstly optimize the problem to find $\hat{x}^{(1)}$, leveraging our inversion attack algorithm while fixing the adapter weight to find $\hat{x}^{(1)}$ that minimizes Equation (\ref{eq:grey_box_problem}); then we fix $\hat{x}^{(1)}$, and only optimize on $\theta_{A_1},\theta_{B_1},\cdots,\theta_{A_{a-1}},\theta_{B_{a-1}}$ LoRA weights to find $\theta_{A_1}^{(2)},\theta_{B_1}^{(2)},\cdots,\theta_{A_{a-1}}^{(2)},\theta_{B_{a-1}}^{(2)}$ that minimizes Equation  ~(\ref{eq:grey_box_problem}). The grey-box inverted result is the last version of $\hat{x}$.  


\begin{table*}[h]
\centering
\scalebox{1.2}{
\begin{tabular}{|l|cc|ccc|cccc|}
\hline
     Participant number        & \multicolumn{2}{c|}{3}     & \multicolumn{3}{c|}{4}                              & \multicolumn{4}{c|}{5}                                                       \\ \hline
Position     & \multicolumn{1}{c|}{2} & 3 & \multicolumn{1}{c|}{2} & \multicolumn{1}{c|}{3} & 4 & \multicolumn{1}{c|}{2} & \multicolumn{1}{c|}{3} & \multicolumn{1}{c|}{4} & 5 \\ \hline
Layer number & \multicolumn{1}{c|}{27}  &  54 & \multicolumn{1}{c|}{20}  & \multicolumn{1}{c|}{40}  &  60 & \multicolumn{1}{c|}{16}  & \multicolumn{1}{c|}{32}  & \multicolumn{1}{c|}{48}  & 64  \\ \hline
Token accuracy     & \multicolumn{1}{c|}{0.9165}  &  0.8896 & \multicolumn{1}{c|}{0.9202}  & \multicolumn{1}{c|}{0.9045}  & 0.8838  & \multicolumn{1}{c|}{0.9320}  & \multicolumn{1}{c|}{0.9098}  & \multicolumn{1}{c|}{0.8987}  & 0.8745  \\ \hline
BLEU         & \multicolumn{1}{c|}{0.8899}  & 0.8492  & \multicolumn{1}{c|}{0.8956}  & \multicolumn{1}{c|}{0.8694}  & 0.8428  & \multicolumn{1}{c|}{0.9145}  & \multicolumn{1}{c|}{0.8803}  & \multicolumn{1}{c|}{0.8573}  &0.8377   \\ \hline
NERR         & \multicolumn{1}{c|}{0.7101}  & 0.6680  & \multicolumn{1}{c|}{0.7533} &  \multicolumn{1}{c|}{0.6881} & 0.6605 & \multicolumn{1}{c|}{0.7730
}  & \multicolumn{1}{c|}{0.6945
}  & \multicolumn{1}{c|}{0.6752
} & 0.6550    \\ 
\hline
\end{tabular}
}
\caption{Evaluation of white-box PIA under different participant number and attacker positions on Skytrax~\cite{skytrax} dataset. }
\label{tab:main_results}

\end{table*}

\begin{table}[!t]
\centering
\scalebox{0.95}{
\begin{tabular}{|l|c|c|c|c|}
\hline
               & \makecell{Song et al. \\ ~\cite{song2020information}} 
               & \makecell{Li et al.\\ ~\cite{Generative-embedding-inversion-attack}}
               & \makecell{Morris et al.\\~\cite{morris2023language} 
               }
               & Ours   \\ \hline
Token accuracy & 0.0040                                                 & 0.2277                                                                 & 0.1476        & \textbf{0.8838} \\ \hline
BLEU           & 0.0021                                                 & 0.1867                                                                 & 0.1293        & \textbf{0.8428} \\ \hline
NERR           & 0                                                      & 0.1098                                                                 & 0.0642        & \textbf{0.6605} \\ \hline
\end{tabular}
}
\caption{Comparison with baselines on Skytrax dataset.}
\vspace{-1em}
\label{fig:baseline}
\end{table}

\begin{table*}[t]
\centering
\scalebox{1}{
\begin{tabular}{|c|c|c|c|c|}
\hline
  & \makecell{Naive optimization \\+ naive discretization} & \makecell{Constrained optimization \\+ naive discretization} & \makecell{Naive optimization \\+ adaptive discretization} & \makecell{\textbf{Constrained optimization} \\\textbf{+ adaptive discretization}} \\ \hline
Token accuracy &       0.3846                                    &                0.5485                                 &         0.6509                                     &                0.8838                                    \\ \hline
BLEU     &        0.2820                                   &        0.4961                                         &     0.5736                                         &    0.8428                                                \\ \hline
NERR     &           0.0399                                &          0.1901                                       &        0.2667                                      &     0.6605                                               \\ \hline
\end{tabular}
}
\caption{Evaluation on different components in our method on Skytrax~\cite{skytrax} dataset.  ``Constrained optimization + adaptive discretization" is our final solution.  }
\label{tab:contribution_results}
\vspace{-1em}
\end{table*}

\section{Experiments}
In this section, we present a comprehensive set of experiments designed to rigorously evaluate the accuracy of our prompt inversion attack for collaborative inference under both white-box and grey-box scenarios. We also dive into the reason why each of our design components works. We provide in-depth analysis to elucidate the rationale behind the functionality of each design component.

\subsection{Setup}
\noindent
\textbf{Participant number and attacker position.} 
In our experiments, we typically assume the involvement of 4 participants for inference on a specific prompt by default, unless stated otherwise. 
Multiple prompts may be served simultaneously by different participant groups. We also explore scenarios with 3 and 5 participants. 
By default, we consider the most challenging scenario for the attacker, where the attacker is positioned as the last participant. In this case, the attacker is required to invert the highest number of layers, adding difficulties to the attack.

\noindent \textbf{Datasets.} 
Following~\cite{pan2020privacy,Sentence_inference_attack}, we utilize 3 datasets from different domains for our evaluation: Skytrax~\cite{skytrax}, CMS~\cite{cms}, and ECHR Law~\cite{echr}. Skytrax is composed of airline reviews from 2006 to 2019 for popular airlines around the world. The CMS dataset records medical information on services and procedures documented by physicians and other healthcare professionals. The ECHR Law dataset contains approximately 11,500 cases recorded in the European Court of Human Rights. Each case in the dataset provides a brief description of legal facts. For each dataset, we evaluate on 150 prompts and compute the average result.



\noindent \textbf{Large language models.}
We invert 3 LLMs in our experiments, including Llama-65B~\cite{llama}, Llama-2-70B~\cite{llama2}, and OPT-66B~\cite{opt}. Llama-65B is an auto-regressive model trained on 1.4T tokens from the 20 languages with the most speakers. Llama-2-70B, trained on a larger training-set (2T tokens), has better context learning ability showed in 4096 token context length. OPT-66B was the largest unrestricted open-sourced pre-trained transformer language model at March 2022, which roughly matched the size and performance of GPT-3 class of models.
We leverage Llama-7B as our oracle model for the semantic-based candidate set generation in our adaptive discretization procedure. 

\noindent \textbf{Metrics. }
We employ three metrics to evaluate how attacks invert the input text, including \textbf{token accuracy}, \textbf{BLEU}~\cite{bleu}, and \textbf{NERR} (named entity recovered ratio)~\cite{nerr}. The first metric measures the ratio of correct tokens at the correct positions in the recovered text. The second metric measures the overlap between the original text and the recovered text, which focuses more on detecting redundant information in recovered text. In addition, we employ the third metric to measure the proportion of keywords (specifically, the key nouns, names, and entities) completely recovered from the model's activations. All three metrics satisfy that higher values indicate inversion results closer to the ground truth, which is better. 


\noindent \textbf{Hyperparameters. }
We set default hyperparameters as optimization learning rate $\beta=0.1$, constraint coefficient $\lambda=0.1$, optimization iteration number $N=2000$, embedding-based candidate size $K=10$, and semantic-based candidate size $Y=10$. For grey-box  attack, the learning rate for LoRA adapter is $10^{-3}$, and the iteration number is 5.



\subsection{Main Results of White-box PIA}

 \begin{figure*}[t]

 \centering
\subfloat[]{\includegraphics[width=0.3\textwidth]{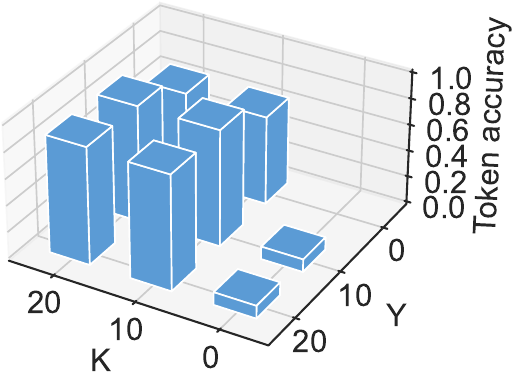}} \hspace{+2mm}
\subfloat[]{\includegraphics[width=0.3\textwidth]{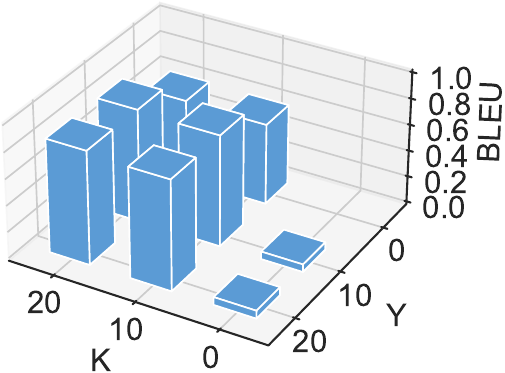}}   \hspace{+2mm}
\subfloat[]{\includegraphics[width=0.3\textwidth]{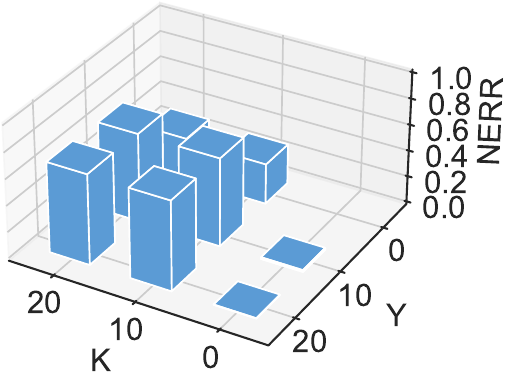}} 

 \caption{Impacts of different $K$ and different $Y$  on Skytrax~\cite{skytrax} dataset.}
\vspace{-1em}
 \label{fig:KY}
 \end{figure*}
 
\noindent
\textbf{Our method achieves high inversion accuracy. } 
To assess the performance of our method under diverse settings, we evaluate all possible attacker positions for different participant numbers 3, 4, and 5 on the Skytrax dataset while keeping other hyper-parameters as default. The results are depicted in Table~\ref{tab:main_results}. We find that token accuracy is consistently high ($>0.87$) across different settings, demonstrating the versatility of our attack. 
For example, in the case of the last participant among 4 participants, the attacker needs to invert 60 layers for the Llama-65B model. Our method achieves a high token accuracy of 0.884 in this scenario. Our method also consistently achieves a high BLEU ($>0.8$) and NERR ($>0.65$). Note that keywords (\eg, geographic name) often consist of multiple tokens, and NERR only counts when the keyword is completely recovered without any error. The effectiveness of the attack marginally diminishes as the attacker is placed in a more backward position. The decline is primarily attributed to the fact that the attacker needs to invert more preceding layers. The non-linearity of the targeted model increases with the layer numbers, making the optimization problem more difficult. In summary, white-box PIA has yielded highly accurate recovery results, even in the most difficult scenarios.

\noindent
\textbf{Our method maintains high efficiency across different models and datasets. }
To showcase the generalizability of our method, 
we conduct experiments across different LLMs and different datasets.
The results are presented in Table~\ref{fig:dataset} and Table~\ref{fig:model} in Appendix. Our method yields similar performance across different models. For example, our attack achieves a high token accuracy on Llama2-70B up to 0.9275. Our method also achieves strong performance on datasets from different domains. On CMS dataset from medical domain and ECHR law dataset, we also achieve token accuracy higher than 0.9. These results indicate our method generalizes well across various settings.

\noindent \textbf{Our method outperforms baselines.} To further evaluate the effectiveness of our approach, we compare with baselines including Song et al. ~\cite{song2020information}, Li et al.~\cite{Generative-embedding-inversion-attack}, and Morris et al.~\cite{morris2023language}. For introduction of these baselines, please refer to Section~\ref{sec:inversion}. We conduct experiments with hyperparameters  under  default settings. The results are shown in Table~\ref{fig:baseline}.  It demonstrates that our method is capable of inverting input prompts with near-perfect accuracy, significantly outperforming existing baselines. 
The major reason for our method outperforming generation-based inversion attacks~\cite{Generative-embedding-inversion-attack,morris2023language} is that, during generation, they only feed the inversion target information to the decoder at the initial phase. When the generation length gets longer, the generated content becomes less related  with the inversion target, the recover error accumulates. Besides, our method outperforms~\cite{song2020information} since our attack does not suffer from gradient vanishing, while our adoption of \textit{constrained optimization} and \textit{adaptive discretization} significantly improves the optimization quality and recovering accuracy in discretization, respectively. 

\noindent\textbf{Impact of our design components.} 
To further understand the effectiveness of individual components of our attack, we also explore replacing the two components of our method with the previously mentioned simple methods (\textit{naive optimization} and \textit{naive discretization}). The results are reported in Table~\ref{tab:contribution_results}. We observe that both two components of our attack are quite crucial and effective. When we adopt the simple methods at both the optimization and discretization phases, the inversion can only achieve 0.3846 in token accuracy. By adopting \textit{constrained optimization} at the optimization phase, the inversion token accuracy improves to 0.5485 (+ 42.6\%). This shows the benefit of constraining $\hat{v}$ to be near $E(\hat{x})$ during the optimization procedure. On the other hand, \textit{adaptive discretization} also greatly boosts the overall inversion performance; it further increases the token accuracy of 0.5485 when only adopting \textit{constrained optimization} to 0.8838, increasing another 61.2\%. This improvement can be attributed to the adaptive discretization policy, the \textit{activation calibration}, and the \textit{semantic speculation}. The results in Table~\ref{tab:main_results} demonstrate that our optimization and discretization are crucial to achieving strong PIA results.

 \begin{figure*}[t]
 \centering

 \subfloat[\label{fig:alpha}]{\includegraphics[width=0.33\textwidth]{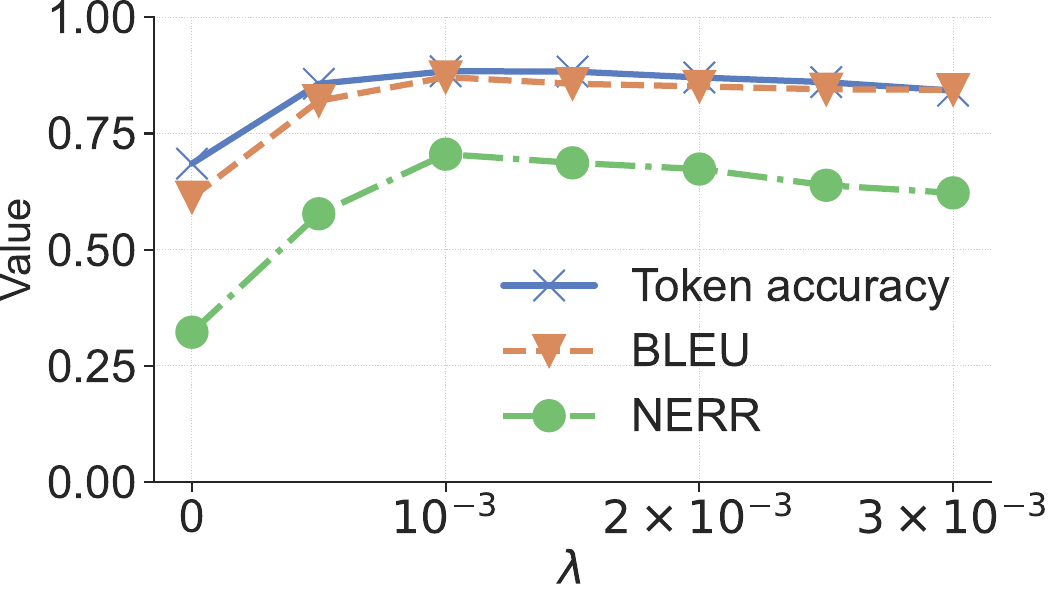}}
  \subfloat[\label{fig:lr}]{\includegraphics[width=0.33\textwidth]{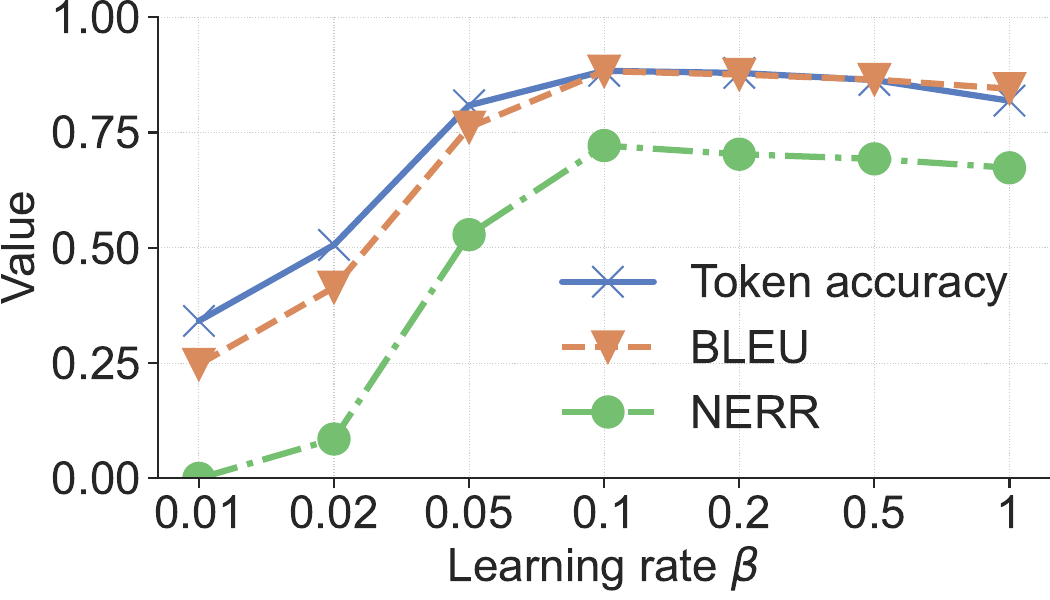}}
  \subfloat[\label{fig:epoch}]{\includegraphics[width=0.33\textwidth]{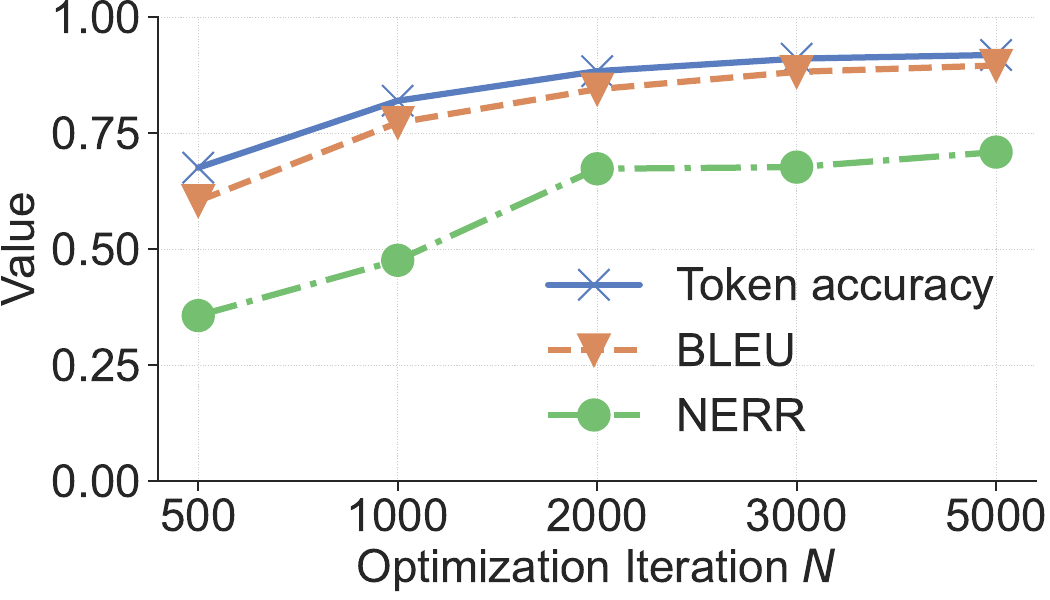}}

 \caption{These figures show the impact of constraint coefficient $\lambda$, learning rate $\beta$, and optimization iteration number $N$ on inversion results, which is tested on Skytrax~\cite{skytrax} dataset.}
 \label{fig:optimization_results}
 \vspace{-1em}
 \end{figure*}




\subsection{Ablation Study}

\noindent\textbf{Impact of constraint  coefficient $\lambda$.} To demonstrate its impact on inversion accuracy, we conduct experiments with $\lambda$ set at $[0,0.0005,0.001,0.0015,0.002,0.0025,0.005,0.01]$, while all other parameters are in default settings. Figure~\ref{fig:alpha}  presents the results of the influence of coefficient $\lambda$. When $\lambda$ increases from 0, the corresponding token accuracy of inversion firstly increases then decreases. We speculate that when $\lambda$ is relatively small, increasing $\lambda$ results in a stronger constraint to $\hat{v}$, improving the optimization result. 
However, when $\lambda$ becomes too large in the initial phase of optimization, the loss becomes overly concentrated on the second constraint term, ignoring the original optimization goal of making the intermediate activations nearer, which causes the optimization stuck in poor local minima.


\noindent\textbf{Impact of learning rate $\beta$.} To demonstrate the impact of learning rate $\beta$ on inversion accuracy, we conduct experiments with $\beta$ ranging from $\{0.01,0.02,0.05,0.1,0.2,0.5,1\}$, while keeping other hyperparameters at their default settings. The results are reported in Figure \ref{fig:lr}. When $\beta$ is small, as $\beta$ increases, the inversion token accuracy increases. We speculate that when $\beta$ is too small, the optimization proceeds too slowly, resulting in sub-optimal inversion with the same number of iterations. Also, when $\beta$ is too large, increasing $\beta$ will cause token accuracy to drop. 
This is because  the optimization oscillates when $\beta$ is excessively large, due to excessively large iteration steps making the convergence difficult.

\noindent\textbf{Impact of optimization iteration number $N$.} To explore the influence of optimization iteration number $N$ on inversion accuracy, we conduct experiments with $N$ ranging from $\{500,1000,2000,3000,5000\}$, while keeping other hyperparameters at their default settings. The results are presented in Figure \ref{fig:epoch}. When $N$ is small, inversion token accuracy significantly increases with the increase of $N$; when $N$ is quite big, increasing $N$ has negligible impact on the inversion token accuracy. This means small iteration number is not enough for the optimization to converge; when $N$ is big enough for the optimization to converge, further optimizing can hardly improve the final inversion result.

\noindent\textbf{Impact of candidate set size $K$ and $Y$.} To demonstrate the impact of candidate set size $K$ and $Y$ on inversion accuracy, we conduct experiments with $K$ and $Y$ both ranging from $\{0,10,20\}$, while keeping other hyperparameters at their default settings. The results are presented in Figure~\ref{fig:KY}. We can draw multiple conclusions from the figure: (1) If we only use the set $S_s$ ($K=0$), the inversion accuracy is quite low. The reason is that only selecting $S_s$ as the candidate set neglects embedding information obtained during optimization; (2) if we only use the set $S_e$ ($Y=0$), the inversion is also sub-optimal. We guess that only leveraging $S_e$ neglects semantic coherence and can not correctly recover under the case of ground-truth not belonging to $S_e$ set.  We present more fine-grained results and analysis on this point in Section~\ref{sec:divein}; 
(3) when $K\ge 10$ or $Y\ge 10$, further increasing $K$ and $Y$ has negligible improvement on the inversion result. It is mostly because the correct token is already in one of the top-10 of $S_e$ and $S_s$ in the most cases, otherwise the token is in the corner case that it does not appear in both two set. In either cases, doubling the candidate size has negligible benefits.



\subsection{Main Results of Grey-box PIA}
\begin{table}[t]
\centering
\scalebox{0.9}{
\begin{tabular}{|l|c|c|c|}
\hline
         & Guanaco (65B) & Medalpaca(30B) & Llama-HH (30B) \\\hline
Token accuracy &     0.8544
        &       0.9049             &    0.8985               \\ \hline
BLEU     &     0.8192
        &     0.8879               &       0.8540            \\ \hline
NERR     &    0.5927
         &     0.6079
              &         0.6152
          \\ \hline
\end{tabular}
}
\caption{Results for grey-box inversion attacks on three LoRA models on Skytrax~\cite{skytrax} dataset.}
\vspace{-2em}
\label{tab:grey}
\end{table}

For exploring the effectiveness of grey-box attack, we conduct experiments on LoRA fine-tuned models including Guanaco-65B~\cite{guanaco}, Medalpaca-lora-30B~\cite{medalpaca}, Llama-HH-lora-30B~\cite{HH}. Here we use some 30B models because the number of open source 65B fine-tuned models is few.  Guanaco models are open-source fine-tuned chat-bots obtained through 4-bit QLoRA~\cite{guanaco} tuning of Llama base models on the OASST1~\cite{oasst1} dataset. MedAlpaca~\cite{medalpaca} are specifically fine-tuned Llama models for medical question-answering and dialogue applications, which are fine-tuned on CMS~\cite{cms} dataset.  Llama-HH is a helpful and harmless natural language assistant fine-tuned on Llama by collecting human preference data and applying the techniques of preference modeling
 and RLHF~\cite{rlhf}. For hyperparameters, we mostly follow the white-box  settings.  For optimization on LoRA weight $\theta_{A_1},\theta_{B_1},\ldots, \theta_{A_{a-1}},\theta_{B_{a-1}}$, we use learning rate of  $10^{-3}$ and take 5 iterations. We also use the default Skytrax dataset for Guanaco and Llama-HH models since they are designed as general chat-bots. For Medalpaca, we use the CMS dataset for evaluation since it is specifically fine-tuned for medical usage.

As shown in Table~\ref{tab:grey}, our grey-box attack achieves satisfying inversion results on different models. Concretely, it achieves over 85\%  token accuracy on all LoRA models under the default setting of inverting 60 layers, which is only marginally lower than that of our white-box method. This verifies the capability of our attack under the more challenging grey-box attack setting.

\section{Potential Defenses}
In this section, we discuss whether our attack is resistant to potential defenses.

\begin{figure}[t]
\centering
\includegraphics[width=0.4\textwidth]{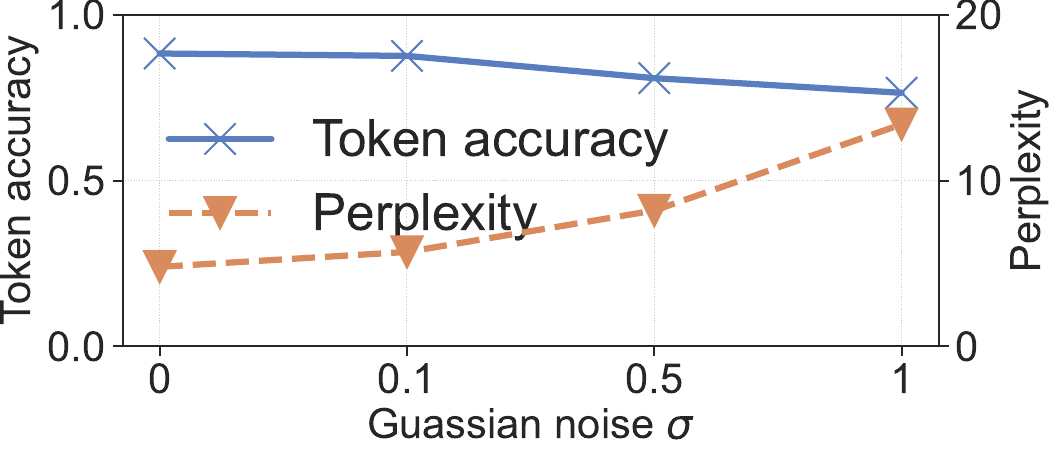}
\caption{Token accuracy and perplexity under defense based on activation obfuscation on Skytrax~\cite{skytrax} dataset. }
\label{fig:DP}
\vspace{-1em}
\end{figure}

\noindent
\textbf{Activation obfuscation. }
Adding random noise to obfuscate sensitive information is a general privacy protection paradigm. 
Without loss of generality, we consider a scheme where each participant adds random Gaussian noise $\epsilon$ to the activation before sending it. We refer to this scheme as \emph{activation obfuscation}. 
We utilize \textit{perplexity}~\cite{perplexity}, measured on a Llama-7B model, to assess the quality of the text generated by the activation obfuscation scheme.  From Figure~\ref{fig:DP} , we can see that when the variance $\sigma$ of noise  $\epsilon$  is small (e.g., 0.1), PIA still recovers most of the prompt information (token accuracy 87.66\%). When noise is greater, the inversion accuracy decreases slightly, while the quality of the generated text is severely reduced. When $\sigma=1$, the perplexity increases to 14.2. These results indicate that \emph{activation obfuscation}  does not achieve a satisfactory trade-off between privacy protection and LLM utility. 


\noindent
\textbf{Activation quantization.}
\emph{Activation quantization} is a widely used technique for minimizing communication in collaborative inference and fine-tuning systems~\cite{pact, cocktail}. 
\revise{Unlike conventional LLM quantization, in activation quantization, the user does not finetune the model weights, as they cannot modify the portions of the weights controlled by other users.  We consider it as a potential defense targetting to reduce information leakage to attackers.} We conduct experiments in the setting where each participant quantizes their activation to 8 bits and 4 bits before sending it. \emph{Activation quantization}  also involves a trade-off between text quality and privacy. From Table~\ref{tab:quantization} in Appendix, we observe that inversion still achieves high accuracy under 8-bit quantization. For 4-bit quantization, the inversion accuracy drops significantly, but the text quality is also severely damaged (perplexity 28.3). These results indicate that \emph{activation quantization} is not an effective defense against our attack.



\noindent
\textbf{Homomorphic encryption.} Leveraging homomorphic encryption~\cite{HE} enables participants to perform DNN computations on encrypted inputs, thereby preserving privacy. However, the computational overhead associated with homomorphically computing an LLM is prohibitively high~\cite{chen-etal-2022-x}, making real-world deployment challenging.


\section{Discussions}

\noindent \textbf{Participant number and model depth.} In our experiments, we do not consider a large number of participants for a single prompt, since involving too many participants for a single prompt is unnecessary and increases communication overhead. For instance, serving a Llama-65B requires only 6 participants each possessing a 24GB GPU. Furthermore, the last participant in 5 participants already needs to invert 80\% layers of the whole LLM. Inverting under the case of 10 participants will not be significantly more difficult. 

\revise{
\noindent \textbf{Ethical considerations.}
We have to emphasize that we do not endorse the practical application of prompt inversion attacks studied. Our objective in presenting this study is to alert the LLM community to the inherent risks of these attacks in collaborative inference scenarios. Accordingly, all experiments in our paper are purely simulated and do not involve attacks on real-world platforms.

We have disclosed the vulnerability studied in this paper to the developers of the PETALS platform and the LocalAI project, granting them a 60-day period to address the issue before publication.
Despite these efforts, we strongly encourage developers to remind users of the risks associated with prompt privacy leakage in collaborative inference and to discourage the input of sensitive information in prompts. Users should also be advised that they can mitigate this risk by collaborating exclusively with trusted participants.

We remain committed to exploring more effective defense mechanisms in future work, although this paper demonstrates that the risk of a prompt inversion attack cannot be simply and directly mitigated using the most classical and common techniques.
}

\noindent \textbf{Future work.} 
Our PIA design philosophies also shed light on improving other forms of inversion attacks on language models. For instance, Our building blocks can greatly enhance the performance of embedding inversion attacks in~\cite{song2020information}. Our method achieves $95.4\%$ F1 score under the setting of ~\cite{song2020information}, while ~\cite{song2020information} only achieves $49.7\%$. 
Additionally, the insights in output calibration of \textit{adaptive discretization} is likely to enhance the performance of the LLM logits inversion~\cite{morris2023language} and LLM output text inversion~\cite{zhang2024extracting}. We will verify them in our future works. 
 Concurrently, Luo et al.~\cite{luo2025prompt} have also investigated a related problem in the black-box setting.

\vspace{-0.3em}
\section{Conclusion}
\vspace{-0.2em}

In this paper, we present a comprehensive exploration of the privacy threats posed by Prompt Inversion Attacks (PIA) in the emerging scenario of LLM collaborative inference. Our novel PIA method marks a significant milestone by nearly perfectly recovering the input prompt from intermediate activations. 
We conduct extensive experiments on benchmark datasets, which verifies the threats of our attack and its resistance to potential defenses,  highlighting vulnerabilities in real-world collaborative LLM inference systems.


\bibliographystyle{IEEEtran}
\bibliography{BIB}
\appendices
\setcounter{section}{0}
\renewcommand{\thesubsection}{\Alph{subsection}} 
\renewcommand{\thesubsubsection}{\thesection.\arabic{subsection}}

\newpage 


\section{Theoretical Analysis of Gradient Vanishing for Softmax Attention Mechanism in \cite{song2020information} }
\label{appendix:ccs20_vanish_theory}
We provide the formal proof of Theorem~\ref{thm:vanish}  as follows. 
\begin{proof}
Since $v=W^\top softmax(z)$, $s=softmax(z)$, 
by chain rule, we have $  \frac{\partial L}{\partial z}= \frac{\partial s}{\partial z}\cdot \frac{\partial v}{\partial s}\cdot \frac{\partial L}{\partial v}$.
 From $v=W^\top s$, we have $\frac{\partial v}{\partial s}={(W^\top)}^\top=W$.
 
By the above equations, considering the $i$-th element in $\frac{\partial L}{\partial z}$, we have $    \frac{\partial L}{\partial z_i}=  \frac{\partial s}{\partial z_i} \cdot W\cdot \frac{\partial L}{\partial v}$.

Classical results on the derivative of softmax~\cite{gao2017properties,elfadel1993softmax} gives $\frac{\partial s_j}{\partial z_i}=s_i (\delta_{ij}-s_j)$, 
where $\delta$ is the Kronecker delta function defined as $
    \delta_{i j}= \begin{cases}0 & \text { if } i \neq j, \\ 1 & \text { if } i=j.\end{cases} $. Based on the above equations, we have:\\
\begin{align}
    \frac{\partial s}{\partial z_i}&=[-s_i s_1, -s_i s_2, \cdots , s_i-s_i^2,-s_i s_{i+1},\cdots,-s_i s_{|V|}] .
\end{align}
Thus we have:\\
\begin{align}
\label{eq:dsdzW}
\begin{aligned}
    (\frac{\partial s}{\partial z_i} \cdot W)_j&=-s_i s_1 W_{1j}-s_i s_2 W_{2j}-\cdots+s_i W_{ij}-s_i^2 W_{ij}-\cdots \\
    &=s_i W_{ij}- \sum_{k=1}^{|V|} s_i s_k W_{kj}.\\
\end{aligned}
\end{align}

We can derive $\frac{\partial L}{\partial z_i}$ from Equation (\ref{eq:dsdzW}):\\
\begin{align}
\label{eq:dLdz}
\begin{aligned}
\frac{\partial L}{\partial z_i}&=\sum_{j=1}^h (\frac{\partial s}{\partial z_i} \cdot W)_j \frac{\partial L}{\partial v_j}=\sum_{j=1}^h \frac{\partial L}{\partial v_j} (s_i W_{ij}- \sum_{k=1}^{|V|} s_i s_k W_{kj})\\
&=s_i\sum_{j=1}^h \frac{\partial L}{\partial v_j} ( W_{ij}- \sum_{k=1}^{|V|}  s_k W_{kj}).
\end{aligned}
\end{align}

By the linear combination of normal distributions:\\
\begin{align}
    W_{ij}- \sum_{k=1}^{|V|}  s_k W_{kj} \sim \mathcal{N}(0,\sigma^2(1-\sum_{k=1}^{|V|} s_k^2)).
\end{align}
Further we have:\\
\begin{align}
\label{eq:further}
    \sum_{j=1}^h \frac{\partial L}{\partial v_j} (W_{ij}- \sum_{k=1}^{|V|}  s_k W_{kj}) \sim \mathcal{N}(0,\sum_{j=1}^h (\frac{\partial L}{\partial v_j})^2 \sigma^2(1-\sum_{k=1}^{|V|} s_k^2)).
\end{align}

 Let $S=\sum_{k=1}^{|V|} s_k^2$. From Equation (\ref{eq:dLdz}) and Equation (\ref{eq:further}), we have:\\
\begin{align}
    \frac{\partial L}{\partial z_i} ~\sim \mathcal{N}(0,s_i^2\sigma^2 (1-S) \sum_{j=1}^h (\frac{\partial L}{\partial v_j})^2).
\end{align}

 Let $\sigma_i= s_i\sigma \sqrt{ (1-S)} \sqrt{\sum_{j=1}^h (\frac{\partial L}{\partial v_j})^2}$. 
 Due to $S\ge 0$, $\sigma_i \le s_i \sigma \sqrt{\sum_{j=1}^h (\frac{\partial L}{\partial v_j})^2}$. Because $\frac{\partial L}{\partial z_i} \sim \mathcal{N}(0,\sigma_i^2)$, by the properies of normal distribution, with high probability, $|\frac{\partial L}{\partial z_i}| \le 4\sigma_i$.   Therefore, with high probability,  we have:\\
\begin{align}
    |\frac{\partial L}{\partial z_i}| \le 4 s_i \sigma \sqrt{\sum_{j=1}^h (\frac{\partial L}{\partial v_j})^2}.
\end{align}

Because $\sum_{i=1}^{|V|}s_i=1$, we have:\\
\begin{align}
||\frac{\partial L}{\partial z} ||_1=\sum_{i=1}^{|V|} |\frac{\partial L}{\partial z_i}| \le 4  \sigma \sqrt{\sum_{j=1}^h (\frac{\partial L}{\partial v_j})^2}.
\end{align}
Therefore we prove Theorem~\ref{thm:vanish}.
\end{proof}

Here we also numerically analyze the case for Llama-7B model. For this model, according to the embedding data distribution, we can set $\sigma=0.025$. We empirically observe that, in most cases, $\sqrt{\sum_{j=1}^h (\frac{\partial L}{\partial v_j})^2}\le 5$. By Theorem~\ref{thm:vanish}, we have $||\frac{\partial L}{\partial z} ||_1\le 0.5$, therefore the average of $|\frac{\partial L}{\partial z_i}|\le \frac{0.5}{|V|}\le 2\times 10^{-5}$. For each $i$ that satisfies $s_i\le 2\times 10^{-4}$, which is 6.25 times average $s_i$, $|\frac{\partial L}{\partial z_i}|\le 10^{-4}$. Therefore, for Llama-7B model, most values in $\frac{\partial L}{\partial z_i}$ are smaller than $10^{-4}$, and their average is smaller than $2\times 10^{-5}$. This theoretical estimation well aligns to the empirical observation of Figure~\ref{fig:zgrad}, where most values in $\frac{\partial L}{\partial z_i}$ belongs to $[10^{-6},10^{-4}]$.

\section{Detailed Algorithms in Our Attack Pipeline}

We elaborate the \textit{constrained optimization} procedure in Algorithm~\ref{alg:optim}, the \textit{adaptive discretization} procedure in Algorithm~\ref{alg:discrete}, and our whole prompt inversion attack algorithm in Algorithm~\ref{alg:core}.
 
\label{appendix:alg}
\begin{algorithm}[!h]
   \caption{\emph{Constrained Optimization}}
   \label{alg:optim}
\begin{algorithmic}
   \STATE {\bfseries Input:} Previous participants' layers compute function $F$;  hidden state size $h$; embedding layer $E$; previous participant sent activation $A$; prompt length $|x|$; optimization learning rate $\beta$; optimization iteration number $N$;   constraining parameter $\lambda$
   \\
     \STATE {\bfseries Output:} Optimized embedding space result $\hat{v}$\\
    \STATE Precompute constraint vector $L,R$\\
    \STATE Define $\hat{x}:\mathbb{N}^{|x|}$
    \STATE Randomly initialize  embedding vector $\hat{v}\in \mathbb{R}^{|x|\times h}$ by uniform distribution.\\
    \FOR{$t=1,2,\cdots, N$}
    \STATE Compute optimization based on Equation (\ref{eq:whole lagrange practical version}):\\
    \STATE $L=\|F(\hat{v})-A\|_2^2+ \lambda \cdot \sum_{i=1}^{|x|} \min_{t \in \mathcal{V}} \|\hat{v}_i -E(t) \|_2^2$\\
    \STATE Back propogate $L$, to calculate gradient: $\frac{\partial L}{\partial \hat{v}}$ \\
    \STATE Update by gradient descent: $\hat{v}=\hat{v}+\beta \cdot \frac{\partial L}{\partial \hat{v}}$ \\
    \STATE Clip $\hat{v}$ using constraint vector $L, R$:\\
    \FOR{$i=1,2,\cdots, |x|$}
    \FOR{$j=1,2,\cdots, h$}
    \STATE $\hat{v}_{ij}=\min (\hat{v}_{ij},L_j)$ \\
    \STATE $\hat{v}_{ij}=\max (\hat{v}_{ij},R_j)$ \\
    \ENDFOR
    \ENDFOR
    \ENDFOR
    \RETURN $\hat{v}$\\
   
\end{algorithmic}
\end{algorithm}

\begin{algorithm}[!h]
   \caption{\emph{Adaptive Discretization}}
   \label{alg:discrete}
\begin{algorithmic}
   \STATE {\bfseries Input:} Optimized embedding vector $\hat{v}$; previous participants' layers compute function $F$; hidden state size $h$; embedding layer $E$; embedding table $e:N\to \mathbb{R}^h$; previous participant activation $A$; prompt length $|x|$;  oracle access to another LLM $\mathsf{OLLM}: N^s\to \mathbb{R}^{|\mathcal{V}|}$, $s$ denotes arbitrary input length. 
   \\
  \STATE {\bfseries Output:} Discretized  result $\hat{x}$ \\
   \FOR{$i=1,2,\cdots, |x|$}
    \STATE Sort $k\in \{1,2,\cdots,|\mathcal{V}|\}$ in an ascending order by  \\
    \STATE $\|\hat{v}_i-e(k)\|_2$ obtaining set $S_1$\\
    \STATE $S_1=\mathsf{Top}-K(S_1)$\\
    \STATE $S_2=\emptyset$\\
    \IF{$i>1$}
    \STATE Predict the token logits using oracle LLM: \\
    \STATE $S=\mathsf{OLLM}([\hat{x}_1,\hat{x}_2,\cdots,\hat{x}_{i-1}])$\\
     \STATE Sort $k\in \{1,2,\cdots,|\mathcal{V}|\}$ in a descending order by $S_k$   obtaining $S_2$\\
     \STATE $S_2=\mathsf{Top}-Y(S_2)$\\
    \ENDIF
    \STATE Define best token candidate $t$, best distance $D$ \\
    \FOR {$j \in S_1\bigcup S_2$}
    \STATE $d=\|F(E([\hat{x}_1,\hat{x}_2,\cdots,\hat{x}_{i-1},j]))-A_i\|_2$\\
    \IF {d<D}
    \STATE $t=j$\\
    \STATE $D=d$\\
    
    \ENDIF
    \ENDFOR
    \STATE $\hat{x}_i=t$\\
    \ENDFOR
    \RETURN $\hat{x}$\\
\end{algorithmic}
\end{algorithm}

\begin{algorithm}[!h]
   \caption{\emph{Prompt Inversion Attack for Collaborative Inference}}
   \label{alg:core}
\begin{algorithmic}
   \STATE {\bfseries Input:} Node number $n$; model layer allocated for each participant $F_1,F_2,\cdots,F_n$; hidden state size $h$; embedding layer $E$; embedding table $e:N\to \mathbb{R}^h$; attacker node ID $a$; previous participant activation $A$; prompt length $|x|$; optimization learning rate $\beta$; optimization iteration number $N$;   constraining parameter $\lambda$; oracle access to another LLM $\mathsf{OLLM}: N^s\to \mathbb{R}^{|\mathcal{V}|}$, $s$ denotes arbitrary input length. 
   \\
   \STATE {\bfseries Output:} Inversion attack  result $\hat{x}$ \\
    \STATE Denote $F=F_1 \circ F_2 \cdots \circ F_{a-1}(\cdot)$\\
    \STATE Perform \textit{constrained optimization} using Algorithm~\ref{alg:optim}:\\
    \STATE $\hat{v}=\mathsf{ConstrainedOptimization}(E,F,A,\beta,N,\lambda)$\\
    \STATE Perform \textit{adaptive discretization} using Algorithm~\ref{alg:discrete}:\\
    \STATE $\hat{x}=\mathsf{AdaptiveDiscretization}(\hat{v},E,e,F,A,\mathsf{OLLM})$\\
    \RETURN $\hat{x}$\\
\end{algorithmic}
\end{algorithm}

\section{Proof of Theorem~\ref{thm:accumulate}}
\label{appendix:error_accumulate}
We provide the formal proof for Theorem~\ref{thm:accumulate} in the following. 
\begin{proof}
By the definition of $L_2$ norm, we have:\\
\begin{align}
\label{eq:discretize error analysis}
\begin{aligned}
    &  \| F(E(\hat{x}))-F(E(x)) \|_2^2 \\
    =& \sum_{i=1}^{|x|} \| F(E(\hat{x}))_i-F(E(x))_i \|_2^2.\\
\end{aligned}
\end{align}

Due to the masked self-attention mechanism adopted by LLMs, the activation of the $i$-th token is determined by itself and all of the previous tokens, unrelated to any of the next tokens. Namely, for $\forall i\in \mathbb{N}^+ ,\forall v \in \mathbb{R}^{i\times h}$, we have:\\
\begin{align}
\label{eq:masked llm analysis}
\begin{aligned}
    &  F(v)_i=  F([v_1,v_2,\cdots,v_i])_i .\\
\end{aligned}
\end{align}

By Equation (\ref{eq:masked llm analysis}), we have:\\
\begin{align}
\label{eq:masked llm analysis 2}
\begin{aligned}
      F(E(\hat{x}))_i &=  F([E(\hat{x_1},\hat{x_2},\cdots,\hat{x_i}])_i \\
      F(E(x))_i &= F([E(x_1,x_2,\cdots,x_i])_i&=[A_1,\cdots,A_j]_i. \\
\end{aligned}
\end{align}

By substituting Equation (\ref{eq:masked llm analysis 2}) into Equation (\ref{eq:discretize error analysis}), we have:\\
\begin{align}
\label{eq:discretize error analysis based on masked llm}
\begin{aligned}
    &  \| F(E(\hat{x}))-F(E(x)) \|_2^2 \\
    =& \sum_{i=1}^{|x|} \| F(E(\hat{x_1},\hat{x_2},\cdots,\hat{x_i}))_i-[A_1,\cdots,A_j]_i \|_2^2.\\
\end{aligned}
\end{align}

From Equation (\ref{eq:discretize error analysis based on masked llm}), for any token $t$, we have: 
\begin{align}
\label{eq:discretize error analysis based on masked llm}
\begin{aligned}
    &  \| F(E(\hat{x}_1,\hat{x}_2,\cdots,\hat{x}_{j-1},t))-[A_1,\cdots,A_j] \|_2^2 \\
    =& \| F(E(\hat{x_1},\hat{x_2},\cdots,\hat{x}_{j-1},t))_j-[A_1,\cdots,A_j]_j \|_2^2 +\\ 
    &\sum_{i=1}^{j-1} \| F(E(\hat{x_1},\hat{x_2},\cdots,\hat{x_i}))_i-[A_1,\cdots,A_j]_i \|_2^2.\\
\end{aligned}
\end{align}

Recall the definition of $\hat{x}_j$:
\begin{align}
\label{eq:best solution for discretize}
\begin{aligned}
\hat{x}_j=\arg\min_{\hat{x}_j\in \mathcal{V}}\| F(E(\hat{x_1},\hat{x_2},\cdots,\hat{x}_{j-1},\hat{x}_j))_j-A_j \|_2^2 .
\end{aligned}
\end{align}

From Equation (\ref{eq:best solution for discretize}), for any token $t$, we have:\\
\begin{align}
\label{eq:it is better than the nearest}
\begin{aligned}
&\| F(E(\hat{x_1},\hat{x_2},\cdots,\hat{x}_{j-1},t))_j-[A_1,\cdots,A_j]_j \|_2^2 \ge\\
&\| F(E(\hat{x_1},\hat{x_2},\cdots,\hat{x}_{j-1},\hat{x}_j))_j-[A_1,\cdots,A_j]_j\|_2^2.
\end{aligned}
\end{align}

Substitute $t=\hat{x}_j$ into Equation (\ref{eq:discretize error analysis based on masked llm}):\\
\begin{align}
\label{eq:discretize error analysis for two cases}
\begin{aligned}
    &  \| F(E(\hat{x}_1,\hat{x}_2,\cdots,\hat{x}_{j-1},\hat{x}_j))-[A_1,\cdots,A_j] \|_2^2 \\
    =& \| F(E(\hat{x_1},\hat{x_2},\cdots,\hat{x}_{j-1},\hat{x}_j))_j-[A_1,\cdots,A_j]_j \|_2^2 +\\ 
    &\sum_{i=1}^{j-1} \| F(E(\hat{x_1},\hat{x_2},\cdots,\hat{x_i}))_i-[A_1,\cdots,A_j]_i \|_2^2.\\
\end{aligned}
\end{align}

By Equation (\ref{eq:it is better than the nearest}) and Equation (\ref{eq:discretize error analysis for two cases}), we have:\\
\begin{align}
\label{eq:discretize error analysis difference}
\begin{aligned}
    &  \| F(E(\hat{x}_1,\hat{x}_2,\cdots,\hat{x}_{j-1},t))-[A_1,\cdots,A_j]) \|_2^2 \\
      -  &  \| F(E(\hat{x}_1,\hat{x}_2,\cdots,\hat{x}_{j-1},\hat{x}_j))-[A_1,\cdots,A_j]) \|_2^2 \\
    =&  \| F(E(\hat{x}_1,\hat{x}_2,\cdots,\hat{x}_{j-1},t))_j-[A_1,\cdots,A_j]_j) \|_2^2 \\
      -  &  \| F(E(\hat{x}_1,\hat{x}_2,\cdots,\hat{x}_{j-1},\hat{x}_j))_j-[A_1,\cdots,A_j]_j) \|_2^2 \\
    \ge& 0.\\
\end{aligned}
\end{align}


Thus we complete the proof of Theorem~\ref{thm:accumulate}.
\end{proof}

\section{More experiment results}
\label{appendix:time}

\begin{table}[ht]
\centering
\scalebox{1}
{
\begin{tabular}{|l|c|c|c|}
\hline
& Skytrax   &  CMS  &  ECHR  \\ \hline
Token accuracy  & 0.8838     & 0.9198
 & 0.9051
 \\ \hline
BLEU & 0.8428
 & 0.9028 & 0.8601 \\ \hline
NERR & 0.6605
 & 0.9667  & 0.9565
 \\ 
 \hline
\end{tabular}
}
\caption{Results on three different datasets.}
\label{fig:dataset}

\end{table}

\begin{table}[ht]
\centering
\scalebox{1}
{
\begin{tabular}{|l|c|c|c|}
\hline
& Llama-65B   &  Llama-2-70B  & OPT-66B  \\ \hline
Token accuracy  & 0.8838     & 0.9275
 & 0.8893
 \\ \hline
BLEU & 0.8428
 & 0.9050 & 0.8456 \\ \hline
NERR & 0.6605
 & 0.7783  & 0.7306
 \\ 
\hline
\end{tabular}
}
\caption{Results on three models on Skytrax~\cite{skytrax} dataset.}
\vspace{-1em}
\label{fig:model}
\end{table}

\begin{table}[ht]
\centering
\begin{tabular}{|l|c|}
\hline                       & Probability (\%) \\ \hline
$\mathsf{Pr}$(Ground-truth  is the nearest token to $\hat{v}_i$)          & 54.9            \\ \hline
$\mathsf{Pr}$(Ground-truth  in $S_e$)          & 69.4            \\ \hline
$\mathsf{Pr}$(Ground-truth  in $S_s$ )       & 30.2            \\ \hline
$\mathsf{Pr}$(Ground-truth  in $S_e\bigcup S_s$)   & 89.5            \\ \hline
\makecell{$\mathsf{Pr}$(Ground-truth is chosen  when belonging to $S_e\bigcup S_s$ ) }& 99.2            \\ \hline
\end{tabular}
\caption{The probability of a ground-truth token  getting into the two candidate sets and their union, respectively. We also report the probability of a ground-truth token being the nearest token to the optimized embedding, and chosen correctly by activation calibration when it is in $S_e\bigcup S_s$. All data are tested on Skytrax dataset. 
}
\label{tab:q2tab1}

\end{table}

\begin{table}[ht]
\centering
\scalebox{1}
{
\begin{tabular}{|l|c|c|c|}
\hline
         & w/o quant & 8 bits & 4 bits \\ \hline
Token accuracy &    0.8838             &   0.7891     &   0.2889     \\ \hline
BLEU     &       0.8428          &   0.7672     &   0.2025     \\ \hline
NERR     &      0.6605      &   0.5094    &       0.0062  \\ \hline
Perplexity     &   4.8          &  10.2      &   28.3     \\ \hline
\end{tabular}
}
\caption{Attack accuracy results with activation quantization based defense on Skytrax~\cite{skytrax} dataset. Perplexity reflects the generated text's quality after quantization.}
\label{tab:quantization}
\end{table}


\section{Meta-Review}

The following meta-review was prepared by the program committee for the 2025
IEEE Symposium on Security and Privacy (S\&P) as part of the review process as
detailed in the call for papers.

\subsection{Summary}
This paper presents a timely study of the privacy risks associated with the collaborative inference of LLMs, where each machine processes a portion of the model and sequentially transmits intermediate activations. The authors introduce the concept of prompt inversion attack (PIA), where an adversary intercepts and analyzes these activations to reconstruct the original input prompt. The attack includes two stages, where the first stage ensures that the recovered embedding remains close to valid token embeddings within the vocabulary of the LLM, and the second stage enhances the attack by using an auxiliary LLM to predict likely next tokens, thus improving the recovery process.

\subsection{Scientific Contributions}
\begin{itemize}
\item Identifies an Impactful Vulnerability
\item Provides a valuable step forward in an established field
\item Establishes a New Research Direction
\end{itemize}

\subsection{Reasons for Acceptance}
\begin{enumerate}
\item This paper presents a novel attack that reveals the privacy risks of the collaborative inferences of LLMs. 
\item The authors conduct extensive evaluations on a wide range of LLMs and obtain convincing results to demonstrate the effectiveness of the attack.
\end{enumerate}



\end{document}